\documentclass[3p]{elsarticle}

\usepackage{epsfig}
\usepackage{amssymb}
\usepackage{amsmath}
\usepackage{amsthm}
\usepackage{subfigure}
\usepackage{xcolor}
\usepackage{float}

\begin{document}
 
\newcommand{\calA}{\mathcal{A}}%
\newcommand{\calB}{\mathcal{B}}%
\newcommand{\calC}{\mathcal{C}}%
\newcommand{\calD}{\mathcal{D}}%
\newcommand{\calE}{\mathcal{E}}%
\newcommand{\calF}{\mathcal{F}}%
\newcommand{\calH}{\mathcal{H}}%
\newcommand{\calI}{\mathcal{I}}%
\newcommand{\calJ}{\mathcal{J}}%
\newcommand{\calK}{\mathcal{K}}%
\newcommand{\calL}{\mathcal{L}}%
\newcommand{\calM}{\mathcal{M}}%
\newcommand{\calN}{\mathcal{N}}%
\newcommand{\calP}{\mathcal{P}}%
\newcommand{\calQ}{\mathcal{Q}}%
\newcommand{\calR}{\mathcal{R}}%
\newcommand{\calS}{\mathcal{S}}%
\newcommand{\calV}{\mathcal{V}}%
\newcommand{\calW}{\mathcal{W}}%
\newcommand{\calX}{\mathcal{X}}%
\newcommand{\calT}{\mathcal{T}}%
\newcommand{\bfa}{{\bf a}}\newcommand{\bfA}{{\bf A}}%
\newcommand{\bfb}{{\bf b}}\newcommand{\bfB}{{\bf B}}%
\newcommand{\bfc}{{\bf c}}\newcommand{\bfC}{{\bf C}}%
\newcommand{\bfd}{{\bf d}}\newcommand{\bfD}{{\bf D}}%
\newcommand{\bfe}{{\bf e}}\newcommand{\bfE}{{\bf E}}%
\newcommand{\bff}{{\bf f}}\newcommand{\bfF}{{\bf F}}%
\newcommand{\bfg}{{\bf g}}\newcommand{\bfG}{{\bf G}}%
\newcommand{\bfh}{{\bf h}}\newcommand{\bfH}{{\bf H}}%
\newcommand{\bfi}{{\bf i}}\newcommand{\bfI}{{\bf I}}%
\newcommand{\bfj}{{\bf j}}\newcommand{\bfJ}{{\bf J}}%
\newcommand{\bfk}{{\bf k}}\newcommand{\bfK}{{\bf K}}%
\newcommand{\bfl}{{\bf l}}\newcommand{\bfL}{{\bf L}}%
\newcommand{\bfm}{{\bf m}}\newcommand{\bfM}{{\bf M}}%
\newcommand{\bfn}{{\bf n}}\newcommand{\bfN}{{\bf N}}%
\newcommand{\bfo}{{\bf o}}\newcommand{\bfO}{{\bf O}}%
\newcommand{\bfp}{{\bf p}}\newcommand{\bfP}{{\bf P}}%
\newcommand{\bfq}{{\bf q}}\newcommand{\bfQ}{{\bf Q}}%
\newcommand{\bfr}{{\bf r}}\newcommand{\bfR}{{\bf R}}%
\newcommand{\bfs}{{\bf s}}\newcommand{\bfS}{{\bf S}}%
\newcommand{\bft}{{\bf t}}\newcommand{\bfT}{{\bf T}}%
\newcommand{\bfu}{{\bf u}}\newcommand{\bfU}{{\bf U}}%
\newcommand{\bfv}{{\bf v}}\newcommand{\bfV}{{\bf V}}%
\newcommand{\bfw}{{\bf w}}\newcommand{\bfW}{{\bf W}}%
\newcommand{\bfx}{{\bf x}}\newcommand{\bfX}{{\bf X}}%
\newcommand{\bfy}{{\bf y}}\newcommand{\bfY}{{\bf Y}}%
\newcommand{\bfz}{{\bf z}}\newcommand{\bfZ}{{\bf Z}}%
\newcommand{\bfell}{\boldsymbol{\ell}}%
\newcommand{\bfimath}{\boldsymbol{\imath}}%
\newcommand{\bfjmath}{\boldsymbol{\jmath}}%
\newcommand{\bfwp}{\boldsymbol{\wp}}%
\newcommand{\bfalpha}{\boldsymbol{\alpha}}%
\newcommand{\bfbeta}{\boldsymbol{\beta}}%
\newcommand{\bfgamma}{\boldsymbol{\gamma}}%
\newcommand{\bfdelta}{\boldsymbol{\delta}}%
\newcommand{\bfepsilon}{\boldsymbol{\epsilon}}%
\newcommand{\bfzeta}{\boldsymbol{\zeta}}%
\newcommand{\bfeta}{\boldsymbol{\eta}}%
\newcommand{\bftheta}{\boldsymbol{\theta}}%
\newcommand{\bfiota}{\boldsymbol{\iota}}%
\newcommand{\bfkappa}{\boldsymbol{\kappa}}%
\newcommand{\bflambda}{\boldsymbol{\lambda}}%
\newcommand{\bfmu}{\boldsymbol{\mu}}%
\newcommand{\bfnu}{\boldsymbol{\nu}}%
\newcommand{\bfxi}{\boldsymbol{\xi}}%
\newcommand{\bfpi}{\boldsymbol{\pi}}%
\newcommand{\bfrho}{\boldsymbol{\rho}}%
\newcommand{\bfsigma}{\boldsymbol{\sigma}}%
\newcommand{\bftau}{\boldsymbol{\tau}}%
\newcommand{\bfupsilon}{\boldsymbol{\upsilon}}%
\newcommand{\bfphi}{\boldsymbol{\phi}}%
\newcommand{\bfchi}{\boldsymbol{\chi}}%
\newcommand{\bfpsi}{\boldsymbol{\psi}}%
\newcommand{\bfomega}{\boldsymbol{\omega}}%
\newcommand{\bfvarepsilon}{\boldsymbol{\varepsilon}}%
\newcommand{\bfvartheta}{\boldsymbol{\vartheta}}%
\newcommand{\bfvarpi}{\boldsymbol{\vaarphi}}%
\newcommand{\bfvarrho}{\boldsymbol{\varrho}}%
\newcommand{\bfvarsigma}{\boldsymbol{\varsigma}}%
\newcommand{\bfvarphi}{\boldsymbol{\varphi}}%
\newfont{\tenbfit}{cmmib10}%
\newfont{\svnbfit}{cmmib8}%
\newfont{\tenbfsl}{cmbxti10}
\newcommand{\bfGamma}{\mbox{\tenbfit\char'00\/}}%
\newcommand{\bfDelta}{\mbox{\tenbfit\char'01\/}}%
\newcommand{\bfTheta}{\boldsymbol{\Theta}}%
\newcommand{\bfLambda}{\boldsymbol{\Lambda}}%
\newcommand{\bfXi}{\mbox{\tenbfit\char'04\/}}%
\newcommand{\bfPi}{\mbox{\tenbfit\char'05\/}}%
\newcommand{\bfSigma}{\boldsymbol{\Sigma}}%
\newcommand{\bfUpsilon}{\mbox{\tenbfit\char'07\/}}%
\newcommand{\bfPhi}{\boldsymbol{\Phi}}
\newcommand{\bfPsi}{\mbox{\tenbfit\char'11\/}}%
\newcommand{\bfOmega}{\boldsymbol{\Omega}}
\newcommand{\sbfGamma}{\mbox{\svnbfit\char'00\/}}%
\newcommand{\sbfDelta}{\mbox{\svnbfit\char'01\/}}%
\newcommand{\sbfTheta}{\mbox{\svnbfit\char'02\/}}%
\newcommand{\sbfLambda}{\mbox{\svnbfit\char'03\/}}%
\newcommand{\sbfXi}{\mbox{\svnbfit\char'04\/}}%
\newcommand{\sbfPi}{\mbox{\svnbfit\char'05\/}}%
\newcommand{\sbfSigma}{\mbox{\svnbfit\char'06\/}}%
\newcommand{\sbfUpsilon}{\mbox{\svnbfit\char'07\/}}%
\newcommand{\sbfPhi}{\mbox{\svnbfit\char'10\/}}%
\newcommand{\sbfPsi}{\mbox{\svnbfit\char'11\/}}%
\newcommand{\sbfOmega}{\mbox{\svnbfit\char'12\/}}%
\newfont{\mmit}{cmmi10}
\newfont{\smit}{cmmi9}
\newfont{\bfMit}{cmmi5}
\newcommand{\iGamma}{\mbox{\mmit\char'00\/}}%
\newcommand{\iDelta}{\mbox{\mmit\char'01\/}}%
\newcommand{\iTheta}{\mbox{\mmit\char'02\/}}%
\newcommand{\iLambda}{\mbox{\mmit\char'03\/}}%
\newcommand{\iXi}{\mbox{\mmit\char'04\/}}%
\newcommand{\iPi}{\mbox{\mmit\char'05\/}}%
\newcommand{\iSigma}{\mbox{\mmit\char'06\/}}%
\newcommand{\iUpsilon}{\mbox{\mmit\char'07\/}}%
\newcommand{\iPhi}{\mbox{\mmit\char'10\/}}%
\newcommand{\iPsi}{\mbox{\mmit\char'11\/}}%
\newcommand{\iOmega}{\mbox{\mmit\char'12\/}}%
\newcommand{\siGamma}{\mbox{\smit\char'00\/}}%
\newcommand{\siDelta}{\mbox{\smit\char'01\/}}%
\newcommand{\siTheta}{\mbox{\smit\char'02\/}}%
\newcommand{\siLambda}{\mbox{\smit\char'03\/}}%
\newcommand{\siXi}{\mbox{\smit\char'04\/}}%
\newcommand{\siPi}{\mbox{\smit\char'05\/}}%
\newcommand{\siSigma}{\mbox{\smit\char'06\/}}%
\newcommand{\siUpsilon}{\mbox{\smit\char'07\/}}%
\newcommand{\siPhi}{\mbox{\smit\char'10\/}}%
\newcommand{\siPsi}{\mbox{\smit\char'11\/}}%
\newcommand{\siOmega}{\mbox{\smit\char'12\/}}%
\newcommand{\ssiGamma}{\mbox{\bfMit\char'00\/}}%
\newcommand{\ssiDelta}{\mbox{\bfMit\char'01\/}}%
\newcommand{\ssiTheta}{\mbox{\bfMit\char'02\/}}%
\newcommand{\ssiLambda}{\mbox{\bfMit\char'03\/}}%
\newcommand{\ssiXi}{\mbox{\bfMit\char'04\/}}%
\newcommand{\ssiPi}{\mbox{\bfMit\char'05\/}}%
\newcommand{\ssiSigma}{\mbox{\bfMit\char'06\/}}%
\newcommand{\ssiUpsilon}{\mbox{\bfMit\char'07\/}}%
\newcommand{\ssiPhi}{\mbox{\bfMit\char'10\/}}%
\newcommand{\ssiPsi}{\mbox{\bfMit\char'11\/}}%
\newcommand{\ssiOmega}{\mbox{\bfMit\char'12\/}}%
\newfont{\tenbbb}{msbm10}%
\newfont{\svnbbb}{msbm8}%
\newcommand{\bbA}{\mbox{\tenbbb A\/}}%
\newcommand{\bbB}{\mbox{\tenbbb B\/}}%
\newcommand{\bbC}{\mbox{\tenbbb C\/}}%
\newcommand{\bbD}{\mbox{\tenbbb D\/}}%
\newcommand{\bbE}{\mbox{\tenbbb E\/}}%
\newcommand{\bbF}{\mbox{\tenbbb F\/}}%
\newcommand{\bbG}{\mbox{\tenbbb G\/}}%
\newcommand{\bbH}{\mbox{\tenbbb H\/}}%
\newcommand{\bbI}{\mbox{\tenbbb I\/}}%
\newcommand{\bbJ}{\mbox{\tenbbb J\/}}%
\newcommand{\bbK}{\mbox{\tenbbb K\/}}%
\newcommand{\bbL}{\mbox{\tenbbb L\/}}%
\newcommand{\bbM}{\mbox{\tenbbb M\/}}%
\newcommand{\bbN}{\mbox{\tenbbb N\/}}%
\newcommand{\bbO}{\mbox{\tenbbb O\/}}%
\newcommand{\bbP}{\mbox{\tenbbb P\/}}%
\newcommand{\bbQ}{\mbox{\tenbbb Q\/}}%
\newcommand{\bbR}{\mbox{\tenbbb R\/}}%
\newcommand{\bbS}{\mbox{\tenbbb S\/}}%
\newcommand{\bbT}{\mbox{\tenbbb T\/}}%
\newcommand{\bbU}{\mbox{\tenbbb U\/}}%
\newcommand{\bbV}{\mbox{\tenbbb V\/}}%
\newcommand{\bbW}{\mbox{\tenbbb W\/}}%
\newcommand{\bbX}{\mbox{\tenbbb X\/}}%
\newcommand{\bbY}{\mbox{\tenbbb Y\/}}%
\newcommand{\bbZ}{\mbox{\tenbbb Z\/}}%
\newcommand{\sbbA}{\mbox{\svnbbb A\/}}%
\newcommand{\sbbB}{\mbox{\svnbbb B\/}}%
\newcommand{\sbbC}{\mbox{\svnbbb C\/}}%
\newcommand{\sbbD}{\mbox{\svnbbb D\/}}%
\newcommand{\sbbE}{\mbox{\svnbbb E\/}}%
\newcommand{\sbbF}{\mbox{\svnbbb F\/}}%
\newcommand{\sbbG}{\mbox{\svnbbb G\/}}%
\newcommand{\sbbH}{\mbox{\svnbbb H\/}}%
\newcommand{\sbbI}{\mbox{\svnbbb I\/}}%
\newcommand{\sbbJ}{\mbox{\svnbbb J\/}}%
\newcommand{\sbbK}{\mbox{\svnbbb K\/}}%
\newcommand{\sbbL}{\mbox{\svnbbb L\/}}%
\newcommand{\sbbM}{\mbox{\svnbbb M\/}}%
\newcommand{\sbbN}{\mbox{\svnbbb N\/}}%
\newcommand{\sbbO}{\mbox{\svnbbb O\/}}%
\newcommand{\sbbP}{\mbox{\svnbbb P\/}}%
\newcommand{\sbbQ}{\mbox{\svnbbb Q\/}}%
\newcommand{\sbbR}{\mbox{\svnbbb R\/}}%
\newcommand{\sbbS}{\mbox{\svnbbb S\/}}%
\newcommand{\sbbT}{\mbox{\svnbbb T\/}}%
\newcommand{\sbbU}{\mbox{\svnbbb U\/}}%
\newcommand{\sbbV}{\mbox{\svnbbb V\/}}%
\newcommand{\sbbW}{\mbox{\svnbbb W\/}}%
\newcommand{\sbbX}{\mbox{\svnbbb X\/}}%
\newcommand{\sbbY}{\mbox{\svnbbb Y\/}}%
\newcommand{\sbbZ}{\mbox{\svnbbb Z\/}}%
\newfont{\tenssit}{cmssqi8 at 10pt}%
\newfont{\svnssit}{cmssqi8 at 7pt}%
\newcommand{\ssa}{\mbox{\tenssit a\/}}%
\newcommand{\ssA}{\mbox{\tenssit A\/}}%
\newcommand{\ssb}{\mbox{\tenssit b\/}}%
\newcommand{\ssB}{\mbox{\tenssit B\/}}%
\newcommand{\ssc}{\mbox{\tenssit c\/}}%
\newcommand{\ssC}{\mbox{\tenssit C\/}}%
\newcommand{\ssd}{\mbox{\tenssit d\/}}%
\newcommand{\ssD}{\mbox{\tenssit D\/}}%
\newcommand{\sse}{\mbox{\tenssit e\/}}%
\newcommand{\ssE}{\mbox{\tenssit E\/}}%
\newcommand{\ssf}{\mbox{\tenssit f\/}}%
\newcommand{\ssF}{\mbox{\tenssit F\/}}%
\newcommand{\ssg}{\mbox{\tenssit g\/}}%
\newcommand{\ssG}{\mbox{\tenssit G\/}}%
\newcommand{\ssh}{\mbox{\tenssit h\/}}%
\newcommand{\ssH}{\mbox{\tenssit H\/}}%
\newcommand{\ssi}{\mbox{\tenssit i\/}}%
\newcommand{\ssI}{\mbox{\tenssit I\/}}%
\newcommand{\ssj}{\mbox{\tenssit j\/}}%
\newcommand{\ssJ}{\mbox{\tenssit J\/}}%
\newcommand{\ssk}{\mbox{\tenssit k\/}}%
\newcommand{\ssK}{\mbox{\tenssit K\/}}%
\newcommand{\ssl}{\mbox{\tenssit l\/}}%
\newcommand{\ssL}{\mbox{\tenssit L\/}}%
\newcommand{\ssm}{\mbox{\tenssit m\/}}%
\newcommand{\ssM}{\mbox{\tenssit M\/}}%
\newcommand{\ssn}{\mbox{\tenssit n\/}}%
\newcommand{\ssN}{\mbox{\tenssit N\/}}%
\newcommand{\sso}{\mbox{\tenssit o\/}}%
\newcommand{\ssO}{\mbox{\tenssit O\/}}%
\newcommand{\ssp}{\mbox{\tenssit p\/}}%
\newcommand{\ssP}{\mbox{\tenssit P\/}}%
\newcommand{\ssq}{\mbox{\tenssit q\/}}%
\newcommand{\ssQ}{\mbox{\tenssit Q\/}}%
\newcommand{\ssr}{\mbox{\tenssit r\/}}%
\newcommand{\ssR}{\mbox{\tenssit R\/}}%
\newcommand{\sss}{\mbox{\tenssit s\/}}%
\newcommand{\ssS}{\mbox{\tenssit S\/}}%
\newcommand{\sst}{\mbox{\tenssit t\/}}%
\newcommand{\ssT}{\mbox{\tenssit T\/}}%
\newcommand{\ssu}{\mbox{\tenssit u\/}}%
\newcommand{\ssU}{\mbox{\tenssit U\/}}%
\newcommand{\ssv}{\mbox{\tenssit v\/}}%
\newcommand{\ssV}{\mbox{\tenssit V\/}}%
\newcommand{\ssw}{\mbox{\tenssit w\/}}%
\newcommand{\ssW}{\mbox{\tenssit W\/}}%
\newcommand{\ssx}{\mbox{\tenssit x\/}}%
\newcommand{\ssX}{\mbox{\tenssit X\/}}%
\newcommand{\ssy}{\mbox{\tenssit y\/}}%
\newcommand{\ssY}{\mbox{\tenssit Y\/}}%
\newcommand{\ssz}{\mbox{\tenssit z\/}}%
\newcommand{\ssZ}{\mbox{\tenssit Z\/}}%
\newcommand{\sssa}{\mbox{\svnssit a\/}}%
\newcommand{\sssA}{\mbox{\svnssit A\/}}%
\newcommand{\sssb}{\mbox{\svnssit b\/}}%
\newcommand{\sssB}{\mbox{\svnssit B\/}}%
\newcommand{\sssc}{\mbox{\svnssit c\/}}%
\newcommand{\sssC}{\mbox{\svnssit C\/}}%
\newcommand{\sssd}{\mbox{\svnssit d\/}}%
\newcommand{\sssD}{\mbox{\svnssit D\/}}%
\newcommand{\ssse}{\mbox{\svnssit e\/}}%
\newcommand{\sssE}{\mbox{\svnssit E\/}}%
\newcommand{\sssf}{\mbox{\svnssit f\/}}%
\newcommand{\sssF}{\mbox{\svnssit F\/}}%
\newcommand{\sssg}{\mbox{\svnssit g\/}}%
\newcommand{\sssG}{\mbox{\svnssit G\/}}%
\newcommand{\sssh}{\mbox{\svnssit h\/}}%
\newcommand{\sssH}{\mbox{\svnssit H\/}}%
\newcommand{\sssi}{\mbox{\svnssit i\/}}%
\newcommand{\sssI}{\mbox{\svnssit I\/}}%
\newcommand{\sssj}{\mbox{\svnssit j\/}}%
\newcommand{\sssJ}{\mbox{\svnssit J\/}}%
\newcommand{\sssk}{\mbox{\svnssit k\/}}%
\newcommand{\sssK}{\mbox{\svnssit K\/}}%
\newcommand{\sssl}{\mbox{\svnssit l\/}}%
\newcommand{\sssL}{\mbox{\svnssit L\/}}%
\newcommand{\sssm}{\mbox{\svnssit m\/}}%
\newcommand{\sssM}{\mbox{\svnssit M\/}}%
\newcommand{\sssn}{\mbox{\svnssit n\/}}%
\newcommand{\sssN}{\mbox{\svnssit N\/}}%
\newcommand{\ssso}{\mbox{\svnssit o\/}}%
\newcommand{\sssO}{\mbox{\svnssit O\/}}%
\newcommand{\sssp}{\mbox{\svnssit p\/}}%
\newcommand{\sssP}{\mbox{\svnssit P\/}}%
\newcommand{\sssq}{\mbox{\svnssit q\/}}%
\newcommand{\sssQ}{\mbox{\svnssit Q\/}}%
\newcommand{\sssr}{\mbox{\svnssit r\/}}%
\newcommand{\sssR}{\mbox{\svnssit R\/}}%
\newcommand{\ssss}{\mbox{\svnssit s\/}}%
\newcommand{\sssS}{\mbox{\svnssit S\/}}%
\newcommand{\ssst}{\mbox{\svnssit t\/}}%
\newcommand{\sssT}{\mbox{\svnssit T\/}}%
\newcommand{\sssu}{\mbox{\svnssit u\/}}%
\newcommand{\sssU}{\mbox{\svnssit U\/}}%
\newcommand{\sssv}{\mbox{\svnssit v\/}}%
\newcommand{\sssV}{\mbox{\svnssit V\/}}%
\newcommand{\sssw}{\mbox{\svnssit w\/}}%
\newcommand{\sssW}{\mbox{\svnssit W\/}}%
\newcommand{\sssx}{\mbox{\svnssit x\/}}%
\newcommand{\sssX}{\mbox{\svnssit X\/}}%
\newcommand{\sssy}{\mbox{\svnssit y\/}}%
\newcommand{\sssY}{\mbox{\svnssit Y\/}}%
\newcommand{\sssz}{\mbox{\svnssit z\/}}%
\newcommand{\sssZ}{\mbox{\svnssit Z\/}}%
\newfont{\gothic}{eufm10}%
\newfont{\sgothic}{eufm7}%
\newcommand{\gta}{\mbox{\gothic a\/}}%
\newcommand{\gtA}{\mbox{\gothic A\/}}%
\newcommand{\gtb}{\mbox{\gothic b\/}}%
\newcommand{\gtB}{\mbox{\gothic B\/}}%
\newcommand{\gtc}{\mbox{\gothic c\/}}%
\newcommand{\gtC}{\mbox{\gothic C\/}}%
\newcommand{\gtd}{\mbox{\gothic d\/}}%
\newcommand{\gtD}{\mbox{\gothic D\/}}%
\newcommand{\gte}{\mbox{\gothic e\/}}%
\newcommand{\gtE}{\mbox{\gothic E\/}}%
\newcommand{\gtf}{\mbox{\gothic f\/}}%
\newcommand{\gtF}{\mbox{\gothic F\/}}%
\newcommand{\gtg}{\mbox{\gothic g\/}}%
\newcommand{\gtG}{\mbox{\gothic G\/}}%
\newcommand{\gth}{\mbox{\gothic h\/}}%
\newcommand{\gtH}{\mbox{\gothic H\/}}%
\newcommand{\gti}{\mbox{\gothic i\/}}%
\newcommand{\gtI}{\mbox{\gothic I\/}}%
\newcommand{\gtj}{\mbox{\gothic j\/}}%
\newcommand{\gtJ}{\mbox{\gothic J\/}}%
\newcommand{\gtk}{\mbox{\gothic k\/}}%
\newcommand{\gtK}{\mbox{\gothic K\/}}%
\newcommand{\gtl}{\mbox{\gothic l\/}}%
\newcommand{\gtL}{\mbox{\gothic L\/}}%
\newcommand{\gtm}{\mbox{\gothic m\/}}%
\newcommand{\gtM}{\mbox{\gothic M\/}}%
\newcommand{\gtn}{\mbox{\gothic n\/}}%
\newcommand{\gtN}{\mbox{\gothic N\/}}%
\newcommand{\gto}{\mbox{\gothic o\/}}%
\newcommand{\gtO}{\mbox{\gothic O\/}}%
\newcommand{\gtp}{\mbox{\gothic p\/}}%
\newcommand{\gtP}{\mbox{\gothic P\/}}%
\newcommand{\gtq}{\mbox{\gothic q\/}}%
\newcommand{\gtQ}{\mbox{\gothic Q\/}}%
\newcommand{\gtr}{\mbox{\gothic r\/}}%
\newcommand{\gtR}{\mbox{\gothic R\/}}%
\newcommand{\gts}{\mbox{\gothic s\/}}%
\newcommand{\gtS}{\mbox{\gothic S\/}}%
\newcommand{\gtt}{\mbox{\gothic t\/}}%
\newcommand{\gtT}{\mbox{\gothic T\/}}%
\newcommand{\gtu}{\mbox{\gothic u\/}}%
\newcommand{\gtU}{\mbox{\gothic U\/}}%
\newcommand{\gtv}{\mbox{\gothic v\/}}%
\newcommand{\gtV}{\mbox{\gothic V\/}}%
\newcommand{\gtw}{\mbox{\gothic w\/}}%
\newcommand{\gtW}{\mbox{\gothic W\/}}%
\newcommand{\gtx}{\mbox{\gothic x\/}}%
\newcommand{\gtX}{\mbox{\gothic X\/}}%
\newcommand{\gty}{\mbox{\gothic y\/}}%
\newcommand{\gtY}{\mbox{\gothic Y\/}}%
\newcommand{\gtz}{\mbox{\gothic z\/}}%
\newcommand{\gtZ}{\mbox{\gothic Z\/}}%
\newcommand{\sgta}{\mbox{\sgothic a\/}}%
\newcommand{\sgtA}{\mbox{\sgothic A\/}}%
\newcommand{\sgtb}{\mbox{\sgothic b\/}}%
\newcommand{\sgtB}{\mbox{\sgothic B\/}}%
\newcommand{\sgtc}{\mbox{\sgothic c\/}}%
\newcommand{\sgtC}{\mbox{\sgothic C\/}}%
\newcommand{\sgtd}{\mbox{\sgothic d\/}}%
\newcommand{\sgtD}{\mbox{\sgothic D\/}}%
\newcommand{\sgte}{\mbox{\sgothic e\/}}%
\newcommand{\sgtE}{\mbox{\sgothic E\/}}%
\newcommand{\sgtf}{\mbox{\sgothic f\/}}%
\newcommand{\sgtF}{\mbox{\sgothic F\/}}%
\newcommand{\sgtg}{\mbox{\sgothic g\/}}%
\newcommand{\sgtG}{\mbox{\sgothic G\/}}%
\newcommand{\sgth}{\mbox{\sgothic h\/}}%
\newcommand{\sgtH}{\mbox{\sgothic H\/}}%
\newcommand{\sgti}{\mbox{\sgothic i\/}}%
\newcommand{\sgtI}{\mbox{\sgothic I\/}}%
\newcommand{\sgtj}{\mbox{\sgothic j\/}}%
\newcommand{\sgtJ}{\mbox{\sgothic J\/}}%
\newcommand{\sgtk}{\mbox{\sgothic k\/}}%
\newcommand{\sgtK}{\mbox{\sgothic K\/}}%
\newcommand{\sgtl}{\mbox{\sgothic l\/}}%
\newcommand{\sgtL}{\mbox{\sgothic L\/}}%
\newcommand{\sgtm}{\mbox{\sgothic m\/}}%
\newcommand{\sgtM}{\mbox{\sgothic M\/}}%
\newcommand{\sgtn}{\mbox{\sgothic n\/}}%
\newcommand{\sgtN}{\mbox{\sgothic N\/}}%
\newcommand{\sgto}{\mbox{\sgothic o\/}}%
\newcommand{\sgtO}{\mbox{\sgothic O\/}}%
\newcommand{\sgtp}{\mbox{\sgothic p\/}}%
\newcommand{\sgtP}{\mbox{\sgothic P\/}}%
\newcommand{\sgtq}{\mbox{\sgothic q\/}}%
\newcommand{\sgtQ}{\mbox{\sgothic Q\/}}%
\newcommand{\sgtr}{\mbox{\sgothic r\/}}%
\newcommand{\sgtR}{\mbox{\sgothic R\/}}%
\newcommand{\sgts}{\mbox{\sgothic s\/}}%
\newcommand{\sgtS}{\mbox{\sgothic S\/}}%
\newcommand{\sgtt}{\mbox{\sgothic t\/}}%
\newcommand{\sgtT}{\mbox{\sgothic T\/}}%
\newcommand{\sgtu}{\mbox{\sgothic u\/}}%
\newcommand{\sgtU}{\mbox{\sgothic U\/}}%
\newcommand{\sgtv}{\mbox{\sgothic v\/}}%
\newcommand{\sgtV}{\mbox{\sgothic V\/}}%
\newcommand{\sgtw}{\mbox{\sgothic w\/}}%
\newcommand{\sgtW}{\mbox{\sgothic W\/}}%
\newcommand{\sgtx}{\mbox{\sgothic x\/}}%
\newcommand{\sgtX}{\mbox{\sgothic X\/}}%
\newcommand{\sgty}{\mbox{\sgothic y\/}}%
\newcommand{\sgtY}{\mbox{\sgothic Y\/}}%
\newcommand{\sgtz}{\mbox{\sgothic z\/}}%
\newcommand{\sgtZ}{\mbox{\sgothic Z\/}}%
\newcommand{\Det}{\hbox{\rm det}\mskip2mu}
\newcommand{\cof}{\hbox{\rm cof}\mskip4mu}
\newcommand{\skw}{\hbox{\rm skw}\mskip3mu}
\newcommand{\sym}{\hbox{\rm sym}\mskip3mu}
\newcommand{\Tr}{\hbox{\rm tr}\mskip2mu}
\newcommand{\Rot}{\hbox{\rm Rot}\mskip2mu}
\newcommand{\Lin}{\hbox{\rm Lin}}
\newcommand{\Orth}{\hbox{\rm Orth}}

\newcommand{\zzcirc}{\raisebox{-0.25ex}{$\mskip1mu\circ$}}
\newcommand{\zcirc}[1]{\overset{\zzcirc}{#1}}
\newcommand{\zztri}{\raisebox{-0.25ex}{$\mskip1mu\scriptscriptstyle\triangle$}}
\newcommand{\ztri}[1]{\overset{\zztri}{#1}}
\newcommand{\zzbox}{\raisebox{-0.05ex}{$\mskip1mu\diamond$}}
\newcommand{\zbox}[1]{\overset{\zzbox}{#1}}

\newcommand{\pards}[2]{\mbox{$\dfrac{\partial #1}{\partial {#2 }}$}}

\newcommand{\vs}{\mskip2mu}
\newcommand{\vvs}{\mskip1mu}
\newcommand{\trans}{\mskip-2mu\scriptscriptstyle\top} 
\newcommand{\sperp}{\scriptscriptstyle\perp\mskip-4mu}
\newcommand{\spar}{\scriptscriptstyle\parallel\mskip-4mu}
\newcommand{\xt}{(\bfx,t)}
\newcommand{\Xt}{(\bfX,t)}
\newcommand{\lj}{[\![}
\newcommand{\rj}{]\!]}
\newcommand{\Blj}{\mbox{$\Big[\kern-0.275em\Big[$}}
\newcommand{\Brj}{\mbox{$\Big]\kern-0.275em\Big]$}}
\newcommand{\zed}{{\bf 0}}
\newcommand{\id}{{\bf 1}}
\newcommand{\onehalf}{\textstyle{\frac{1}{2}}}
\newcommand{\third}{\textstyle{\frac{1}{3}}}
\newcommand{\Grad}{\hbox{\rm grad}\mskip2mu}
\newcommand{\Div}{\hbox{\rm Div}\mskip2mu}                                      
\newcommand{\curl}{\hbox{\rm curl}\mskip2mu}         
\newcommand{\Curl}{\hbox{\rm Curl}\mskip2mu}
\newcommand{\divx}{\text{div}\vvs}                   
\newcommand{\B}{\text{B}}
\newcommand{\Bt}{\calB_t}                                                                                    
\newcommand{\dBt}{\partial\calB_t}                                                                          
\newcommand{\p}{\text{P}}                                                                                    
\newcommand{\pt}{\calP_t}
\newcommand{\dpt}{\partial\calP_t}
\newcommand{\X}{\bfX}
\newcommand{\x}{\bfx}
\newcommand{\y}{\bfchi}
\newcommand{\F}{\bfF}
\newcommand{\U}{\bfU}
\newcommand{\V}{\bfV}
\newcommand{\C}{\bfC}
\newcommand{\R}{\bfR}                                                                                              
\newcommand{\FT}{\bfF^{\trans}}                                                                         
\newcommand{\Q}{\bfQ}                                                                                        
\newcommand{\QT}{\bfQ^{\trans}}
\newcommand{\W}{\bfW}
\newcommand{\D}{\bfD}
\newcommand{\A}{\bfA}

\newcommand{\tendot}{\mskip-3mu:\mskip-2mu}
\newcommand{\Def}{\overset{\text{def}}{=}}
\newcommand{\n}{\bfn}
\newcommand{\ep}{\varepsilon}%
\newcommand{\thet}{\vartheta}%

\newcommand{\scA}{{\scriptscriptstyle\sc A}}%
\newcommand{\scB}{{\scriptscriptstyle\sc B}}%
\newcommand{\scC}{{\scriptscriptstyle\sc C}}%
\newcommand{\scD}{{\scriptscriptstyle\sc D}}%
\newcommand{\scE}{{\scriptscriptstyle\sc E}}%
\newcommand{\scF}{{\scriptscriptstyle\sc F}}%
\newcommand{\scG}{{\scriptscriptstyle\sc G}}%
\newcommand{\scH}{{\scriptscriptstyle\sc H}}%
\newcommand{\scI}{{\scriptscriptstyle\sc I}}%
\newcommand{\scJ}{{\scriptscriptstyle\sc J}}%
\newcommand{\scK}{{\scriptscriptstyle\sc K}}%
\newcommand{\scM}{{\scriptscriptstyle\sc M}}%
\newcommand{\scR}{{\scriptscriptstyle\sc R}}%
\newcommand{\scS}{{\scriptscriptstyle\sc S}}%

\newcommand{\mat}{\text{\tiny R}}%
\newcommand{\spat}{\text{\tiny S}}%

\newcommand{\ddt}{\frac{\text{d}}{\text{d}t}}
\newcommand{\dt}[1]{\frac{\text{d}#1}{\text{d}t}}
\renewcommand{\v}{\bfv}%
\renewcommand{\cof}{\scriptscriptstyle\sf{C}}

\newcommand{\rdot}{\dot{\mskip-3mu\phantom{p}}}

\newcommand{\RT}{\bfR^{\trans}}



\newcommand{\boltz}{\text{\tiny B}}%
\newcommand{\hen}{\text{\tiny H}}%
\newcommand{\calG}{\mathcal{G}}%
\newcommand{\lock}{\text{\tiny L}}%

\newcommand{\Ab}{\mbox{\boldmath ${{\bf {\cal A}}}$}}
\newcommand{\Bb}{\mbox{\boldmath ${{\bf {\cal B}}}$}}
\newcommand{\Cb}{\mbox{\boldmath ${{\bf {\cal C}}}$}}
\newcommand{\Db}{\mbox{\boldmath ${{\bf {\cal D}}}$}}
\newcommand{\Eb}{\mbox{\boldmath ${{\bf {\cal E}}}$}}
\newcommand{\Fb}{\mbox{\boldmath ${{\bf {\cal F}}}$}}
\newcommand{\Gb}{\mbox{\boldmath ${{\bf {\cal G}}}$}}
\newcommand{\Hb}{\mbox{\boldmath ${{\bf {\cal H}}}$}}
\newcommand{\Ib}{\mbox{\boldmath ${{\bf {\cal I}}}$}}
\newcommand{\Ibt}{\mbox{\boldmath $\tilde{{\bf {\cal I}}}$}}
\newcommand{\Jb}{\mbox{\boldmath ${{\bf {\cal J}}}$}}
\newcommand{\Kb}{\mbox{\boldmath ${{\bf {\cal K}}}$}}
\newcommand{\Lb}{\mbox{\boldmath ${{\bf {\cal L}}}$}}
\newcommand{\Mb}{\mbox{\boldmath ${{\bf {\cal M}}}$}}
\newcommand{\Nb}{\mbox{\boldmath ${{\bf {\cal N}}}$}}
\newcommand{\Ob}{\mbox{\boldmath ${{\bf {\cal O}}}$}}
\newcommand{\Pb}{\mbox{\boldmath ${{\bf {\cal P}}}$}}
\newcommand{\Qb}{\mbox{\boldmath ${{\bf {\cal Q}}}$}}
\newcommand{\Rb}{\mbox{\boldmath ${{\bf {\cal R}}}$}}
\newcommand{\Tb}{\mbox{\boldmath ${{\bf {\cal T}}}$}}
\newcommand{\Ub}{\mbox{\boldmath ${{\bf {\cal U}}}$}}
\newcommand{\Vb}{\mbox{\boldmath ${{\bf {\cal V}}}$}}
\newcommand{\Wb}{\mbox{\boldmath ${{\bf {\cal W}}}$}}
\newcommand{\Xb}{\mbox{\boldmath ${{\bf {\cal X}}}$}}
\newcommand{\Yb}{\mbox{\boldmath ${{\bf {\cal Y}}}$}}
\newcommand{\Zb}{\mbox{\boldmath ${{\bf {\cal Z}}}$}}
\newcommand{\calU}{\mathcal{U}}%

\newcommand{\thA}{\mbox{$\Bbb{A}$}}
\newcommand{\thB}{\mbox{$\Bbb{B}$}}
\newcommand{\thC}{\mbox{$\Bbb{C}$}}
\newcommand{\thD}{\mbox{$\Bbb{D}$}}
\newcommand{\thE}{\mbox{$\Bbb{E}$}}
\newcommand{\thF}{\mbox{$\Bbb{F}$}}
\newcommand{\thG}{\mbox{$\Bbb{G}$}}
\newcommand{\thH}{\mbox{$\Bbb{H}$}}
\newcommand{\thI}{\mbox{$\Bbb{I}$}}
\newcommand{\thJ}{\mbox{$\Bbb{J}$}}
\newcommand{\thK}{\mbox{$\Bbb{K}$}}
\newcommand{\thL}{\mbox{$\Bbb{L}$}}
\newcommand{\thM}{\mbox{$\Bbb{M}$}}
\newcommand{\thN}{\mbox{$\Bbb{N}$}}
\newcommand{\thO}{\mbox{$\Bbb{O}$}}
\newcommand{\thP}{\mbox{$\Bbb{P}$}}
\newcommand{\thQ}{\mbox{$\Bbb{Q}$}}
\newcommand{\thR}{\mbox{$\Bbb{R}$}}
\newcommand{\thS}{\mbox{$\Bbb{S}$}}
\newcommand{\thT}{\mbox{$\Bbb{T}$}}
\newcommand{\thU}{\mbox{$\Bbb{U}$}}
\newcommand{\thV}{\mbox{$\Bbb{V}$}}
\newcommand{\thW}{\mbox{$\Bbb{W}$}}
\newcommand{\thX}{\mbox{$\Bbb{X}$}}
\newcommand{\thY}{\mbox{$\Bbb{Y}$}}
\newcommand{\thZ}{\mbox{$\Bbb{Z}$}}

\newcommand{\sdiv}{\hbox{\rm div}\mskip2mu}
\newcommand{\scurl}{\hbox{\rm curl}\mskip2mu}
\newcommand{\grad}{\text{grad}\,}

\newcommand{\pardd}[2]{\mbox{${\dfrac{\partial^2}{\partial {#1} \partial{#2}}}$}}
\newcommand{\pardt}[2]{\mbox{${\dfrac{\partial }{\partial {#2}  } {#1} }$}}
\newcommand{\threedot}{\mskip1mu\raisebox{0.1ex}.\mskip-5mu\raisebox{0.6ex}.\mskip-5mu\raisebox{1.1ex}.\mskip1mu}
\newcommand{\somega}{\boldsymbol{\omega}}%
\newcommand{\kappab}{\boldsymbol{\kappa}}%
\newcommand{\kappabb}{\bar{\boldsymbol{\kappa}}}%
\newcommand{\T}{\bfT}
\newcommand{\Te}{\bfT^e}
\newcommand{\Me}{\bfM^e}
\newcommand{\Tealpha}{\bfT^{e\,(\alpha)}}
\newcommand{\Talpha}{\bfT^{(\alpha)}}
\newcommand{\Tv}{\bfT^v}
\newcommand{\Tp}{\bfT^p}
\newcommand{\Tpk}{\bfT_{\text{2pk}}}
\newcommand{\pk}{\text{\tiny PK}}%
\newcommand{\kk}{\text{\tiny K}}%

\newcommand{\Tmat}{\bfT_{\mat}}
\newcommand{\mvs}{\mskip-2mu}

\newcommand{\h}{\bfh}
\newcommand{\DelX}{\Delta\X}
\newcommand{\Delx}{\Delta\x}
\newcommand{\Delv}{\Delta v}
\newcommand{\DelvR}{\Delta v_{\mat}}
\newcommand{\Dela}{\Delta a}
\newcommand{\DelaR}{\Delta a_{\mat}}
\newcommand{\subt}{_{(t)}}
\newcommand{\pullT}{\bbP}
\newcommand{\pullN}{\underline{\bbP}}
\newcommand{\pullTi}{\bbP^{-1}}
\newcommand{\pullNi}{\underline{\bbP}^{-1}}
\newcommand{\vecpullT}{\bfP}
\newcommand{\vecpullN}{\underline{\bfP}}
\newcommand{\vecpullTi}{\bfP^{-1}}
\newcommand{\vecpullNi}{\underline{\bfP}^{-1}}
\newcommand{\ts}{\bft_{\mskip-2mu\scriptscriptstyle\calS}}
\newcommand{\SR}{\text{S}}
\newcommand{\SO}{\calS}
\newcommand{\SRt}{\SR(t)}
\newcommand{\SOt}{\calS(t)}
\newcommand{\VR}{V_{\mat}}
\newcommand{\mR}{\bfm_{\mat}}
\newcommand{\PR}{\bfP_{\mskip-3mu\mat}}
\newcommand{\doty}{\dot{\y}}
\newcommand{\zp}[1]{{#1}^{\scriptscriptstyle+}}
\newcommand{\zm}[1]{{#1}^{\scriptscriptstyle-}}
\newcommand{\zpm}[1]{{#1}^{\scriptscriptstyle\pm}}
\newcommand{\tR}{\text{R}}
\newcommand{\IB}{\calI_{\bfB}}
\newcommand{\Iid}{\calI_{\bf1}}
\newcommand{\IBt}{\tilde{\calI}_{\bfB}}
\newcommand{\jdR}{\jmath_{\scriptscriptstyle\partial\mskip-2mu R}}
\newcommand{\negzeta}{\scriptscriptstyle\neg\mskip-9.75mu\zeta}
\newcommand{\calY}{\mathcal{Y}}%
\newcommand{\nsites}{n^{\text{sites}}}

\newcommand{\cond}{\text{\small K}}%
\newcommand{\enth}{\text{\small H}}%
\newcommand{\gibb}{\text{\small G}}%
\newcommand{\peq}{ p_{\rm eq} }%

\newcommand{\Sym}{\text{sym}}
\newcommand{\Symz}{\text{sym}_0}
\newcommand{\Fe}{\bfF^e}
\newcommand{\Fetilde}{\tilde{\bfF}^{e}}
\newcommand{\Fp}{\bfF^p}
\newcommand{\Fei}{\bfF^{e-1}}
\newcommand{\Fet}{\bfF^{e\,\trans}}
\newcommand{\Feit}{\bfF^{e-\trans}}
\newcommand{\Fpi}{\bfF^{p-1}}
\newcommand{\Le}{\bfL^e}
\newcommand{\Lp}{\bfL^p}
\newcommand{\Dp}{\bfD^p}
\newcommand{\Dptilde}{\tilde{\bfD}^{p}}
\newcommand{\Wp}{\bfW^p}           
\newcommand{\De}{\bfD^e}           
\newcommand{\We}{\bfW^e}
\newcommand{\Ed}{\bfE^d}
\newcommand{\Ee}{\bfE^e}           
\newcommand{\Ep}{\bfE^p}
\newcommand{\He}{\bfH^e}           
\newcommand{\Hp}{\bfH^p}
\newcommand{\dev}{\text{dev}}
\newcommand{\Ce}{\bfC^e}
\newcommand{\Cedot}{\dot{\bfC}^e}
\newcommand{\Eedot}{\dot{\bfE}^e}
\newcommand{\Epdot}{\dot{\bfE}^p}
\newcommand{\Epdoto}{\dot{\bfE}^p_0}
\newcommand{\Np}{\bfN^p}
\newcommand{\Mback}{\bfM_{\text{back}}}
\newcommand{\Meff}{\Me_{\text{eff}}}
\newcommand{\Medot}{\dot{\bfM}{}^e}
\newcommand{\prestar}{{}_*\mskip-2mu}
\newcommand{\epdot}{\dot{\epsilon}^{p} }
\newcommand{\ebarp}{\mbox{$\varepsilon^p$}}
\newcommand{\ebarpdot}{\mbox{$\dot{\varepsilon}{}^p$}}
\newcommand{\nup}{\mbox{$\nu^p$}}
\newcommand{\es}{\mbox{$S$}}
\newcommand{\cee}{\mbox{$C$}}
\newcommand{\hh}{\mbox{$h$}}
\newcommand{\Tback}{\mbox{$\bfT_{\text{back}}$}}
\newcommand{\To}{\mbox{$\bfT_0$}}
\newcommand{\taub}{\mbox{$\bar\tau$}}
\newcommand{\pbar}{\mbox{$\bar{p}$}}
\newcommand{\gammab}{\mbox{$\bar\gamma$}}
\newcommand{\etab}{\mbox{$\bar\eta$}}
\newcommand{\matt}{\text{\tiny M}}%
\newcommand{\rel}{\text{\tiny I}}%
\newcommand{\Trel}{\T_{\rel\rel}}
\newcommand{\ICe}{\calI_{\Ce}}
\newcommand{\Cdis}{\C_{\dis} }
\newcommand{\ICdis}{\calI_{\Cdis}}
\newcommand{\IA}{\calI_{\A}}
\newcommand{\IC}{\calI_{\C}}
\newcommand{\Ue}{\bfU^e}
\newcommand{\Ve}{\bfV^e}
\newcommand{\Up}{\bfU^p}
\newcommand{\Vp}{\bfV^p}
\newcommand{\Rp}{\bfR^p}
\newcommand{\Be}{\bfB^e}
\newcommand{\Bp}{\bfB^p}
\newcommand{\Yp}{\bfY^p}
\newcommand{\Cp}{\bfC^p}
\newcommand{\rot}{\text{Orth}^+}
\newcommand{\ski}{\mbox{\boldmath$\xi$}}
\newcommand{\Imeff}{\calI_{\Meff}}

\newcommand{\Fealpha}{\bfF^{e\,(\alpha)}}
\newcommand{\Fealphat}{\tilde{\bfF}^{e\,(\alpha)}}
\newcommand{\Fedotalpha}{\dot{\bfF}^{e\,(\alpha)}}
\newcommand{\Fedot}{\dot{\bfF}^{e}}
\newcommand{\Realpha}{\bfR^{e\,(\alpha)}}
\newcommand{\Uealpha}{\bfU^{e\,(\alpha)}}
\newcommand{\Vealpha}{\bfV^{e\,(\alpha)}}
\newcommand{\Aalpha}{\bfA^{(\alpha)}}
\newcommand{\Bealpha}{\bfB^{e\,(\alpha)}}
\newcommand{\Cealpha}{\bfC^{e\,(\alpha)}}
\newcommand{\Cedotalpha}{\dot{\bfC}^{e\,(\alpha)}}
\newcommand{\Fpalpha}{\bfF^{p\,(\alpha)}}
\newcommand{\Lpalpha}{\bfL^{p\,(\alpha)}}
\newcommand{\Mpalpha}{\bfM^{p\,(\alpha)}}
\newcommand{\Dpalphat}{\tilde{\bfD}^{p\,(\alpha)}}
\newcommand{\Dpalpha}{\bfD^{p\,(\alpha)}}
\newcommand{\Dpalphab}{\bar{\bfD}^{p\,(\alpha)}}
\newcommand{\Npalpha}{\bfN^{p\,(\alpha)}}
\newcommand{\lpalpha}{l^{p\,(\alpha)}}
\newcommand{\dpalpha}{d^{p\,(\alpha)}}
\newcommand{\deep}{d^{p}}
\newcommand{\Npalphab}{\bar{\bfN}^{p\,(\alpha)}}
\newcommand{\nupalpha}{d^{p\,(\alpha)}}
\newcommand{\nupalphab}{\bar{d}^{p\,(\alpha)}}
\newcommand{\Wpalpha}{\bfW^{p\,(\alpha)}}
\newcommand{\Upalpha}{\bfU^{p\,(\alpha)}}
\newcommand{\Cpalpha}{\bfC^{p\,(\alpha)}}
\newcommand{\Vpalpha}{\bfV^{p\,(\alpha)}}
\newcommand{\Bpalpha}{\bfB^{p\,(\alpha)}}
\newcommand{\Rpalpha}{\bfR^{p\,(\alpha)}}
\newcommand{\Ypalpha}{\bfY^{p\,(\alpha)}}
\newcommand{\Lealpha}{\bfL^{e\,(\alpha)}}
\newcommand{\Dealpha}{\bfD^{e\,(\alpha)}}
\newcommand{\Eealpha}{\bfE^{e\,(\alpha)}}
\newcommand{\Wealpha}{\bfW^{e\,(\alpha)}}
\newcommand{\Qalpha}{\bfQ^{(\alpha)}}
\newcommand{\Qalphat}{\bfQ^{(\alpha)\,\trans}}
\newcommand{\Lealphat}{\tilde{\bfL}^{e\,(\alpha)}}
\newcommand{\Salpha}{S^{(\alpha)}}
\newcommand{\skialpha}{\bfxi^{(\alpha)}}
\newcommand{\xialpha}{\xi^{(\alpha)}}
\newcommand{\halpha}{h^{(\alpha)}}
\newcommand{\ealpha}{\psi^{(\alpha)}}
\newcommand{\lalpha}{\mbox{$\bar{\lambda}^{e\,(\alpha)} $}}
\newcommand{\Pealpha}{\bfP^{e\,(\alpha)}}
\newcommand{\Pe}{\bfP^e}
\newcommand{\Sealpha}{\bfS^{e\,(\alpha)}}
\newcommand{\Mealpha}{\bfM^{e\,(\alpha)}}
 \newcommand{\Jealpha}{J^{e\,(\alpha)}}
\newcommand{\Yalpha}{Y^{(\alpha)}}
\newcommand{\npalpha}{\nu^{p\,(\alpha)}}
\newcommand{\Sigmae}{\boldsymbol{\Sigma}^{e\,(\alpha)}}
\newcommand{\bfSigmae}{\boldsymbol{\Sigma}^{e}}
\newcommand{\Sigmaeo}{\boldsymbol{\Sigma}^{e}_0}
\newcommand{\Sigmaealpha}{\boldsymbol{\Sigma}^{e\,(\alpha)}}
\newcommand{\Sigmaealphao}{\boldsymbol{\Sigma}^{e\,(\alpha)}_0}
\newcommand{\Mealphao}{\bfM^{e\,(\alpha)}_0}
\newcommand{\taup}{\bar{\tau}_p}
\newcommand{\prt}{\text{P}}
\newcommand{\symz}{\text{sym}_0}
\newcommand{\tauact}{ \tau_{\text{e}}}
\newcommand{\dis}{\text{dis}}
\newcommand{\vol}{\text{vol}}
\newcommand{\maxm}{\text{max}}
\newcommand{\refn}{\text{ref}}
\newcommand{\dee}{\mbox{$d$}}

\newcommand{\dX}{\bfd\bfX}
\newcommand{\dx}{\bfd\bfx}
\newcommand{\lX}{\bfu_{\text{R}}}
\newcommand{\lx}{\bar{\bfu}}
\newcommand{\MX}{\calM_{\bfX}}
\newcommand{\refX}{\text{Ref}_{\bfX}}
\newcommand{\defx}{\text{Def}_{\bfx}}
\newcommand{\nX}{\bfn_{\text{R}}}
\newcommand{\nx}{\bar{\bfn}}
\newcommand{\Tpalpha}{\bfT^{p\,(\alpha)}}
\newcommand{\palpha}{\psi^{(\alpha)}}
\newcommand{\Se}{\bfS^e}
\newcommand{\Sv}{\bfS^v}
\newcommand{\Sp}{\bfS^p}
\newcommand{\Spalpha}{\bfS^{p\,(\alpha)}}
\newcommand{\eSalpha}{S^{(\alpha)}}
\newcommand{\Tpdis}{\hat{\bfT}^p_{\text{dis}}}
\newcommand{\Tpen}{\hat{\bfT}^p_{\text{en}}}
\newcommand{\TX}{\bfT(\bfX)}

\newcommand{\lamp}{\bar\lambda^p}
\newcommand{\sign}{\text{sign}}

\begin{frontmatter}

\title{Elastic Interaction of Pressurized Cavities in Hyperelastic Media: Attraction and Repulsion}
\author[inst1]{Ali Saeedi}
\author[inst1]{Mrityunjay Kothari\corref{cor1}}
\address[inst1]{Department of Mechanical Engineering, College of Engineering and Physical Sciences, University of New Hampshire, Durham, NH 03824, USA}

\cortext[cor1]{Corresponding Author. Email: Mrityunjay.Kothari@unh.edu }

\begin{abstract}
This study computationally investigates the elastic interaction of two pressurized cylindrical cavities in a 2D hyperelastic medium. 
Unlike linear elasticity, where interactions are exclusively attractive, nonlinear material models (neo-Hookean, Mooney-Rivlin, Arruda-Boyce) exhibit both attraction and repulsion between the cavities. 
A critical pressure-shear modulus ratio governs the transition, offering a pathway to manipulate cavity configurations through material and loading parameters.
At low ratios, the interactions are always attractive; at higher ratios, both attractive and repulsive regimes exist depending on the separation between the cavities.
Effect of strain stiffening on these interactions are also analyzed.
These insights bridge theoretical and applied mechanics, with implications for soft material design and subsurface engineering.

\keyword{
Interaction Energy \sep Cavity Pair \sep Pressurized Cavity \sep Hyperelastic Material \sep Driving Force}
\endkeyword
\end{abstract}

\end{frontmatter}

\makeatother


\section{Introduction and Background}
\label{sec:intro}

Elastic interactions between embedded pressurized cavities are fundamental to a diverse array of applications, from material science to geomechanics, influencing phenomena like bubble migration and coalescence, the mechanical stability of porous media, phase separation in soft solids, and structural stability in geomechanics \citep{Kothari2020, Kothari2023crucial, Kooi2001, barney2020cavitation, jeffries2011he, zhang2019simulation, xu2019gravity}.
Understanding the mechanics of such interactions is essential for predicting material behavior under complex loading conditions and for tailoring designs in cutting-edge engineering systems.

Classical studies of pressurized cavities in linear elastic media have provided significant insights into elastic interaction effects.
Early works typically relied on solving the governing biharmonic equation from stress function approaches, followed by determining the unknown coefficients from boundary conditions.  
Jeffrey \citep{jeffery1921ix} pioneered the use of bipolar coordinates to address configurations such as two cavities in an infinite plate, a circular cavity near a straight edge, and an eccentric circular cavity in a disk.  
Howland and Knight \citep{howland1939stress} extended these methods to more complex geometries, including infinite rows of traction-free cavities subjected to external stresses.  
Green \citep{Green1940} advanced the field by formulating a general solution for plates containing traction-free cavities of varying sizes, providing numerical results for cases involving three cavities.  
Ling \citep{Ling1948} made an important contribution by deriving explicit series solutions for stresses around two equal, traction-free circular cavities under remote uniaxial, transverse, and equibiaxial loads.  
Iwaki and Miyao \citep{toshihiro1980stress} and Hoang and Abousleiman \citep{Hoang2008} addressed the problem of internally pressurized cavities in infinite plates subjected to remote stresses.  
These studies concentrated on examining the resultant stress fields and their characteristics, including their peak values; they did not explore the energetics of the problem.
A pivotal but flawed contribution came from Davanas \citep{Davanas1992}, who computed the elastic interaction energy of equal, pressurized cavities in an isotropic material from the stress and strain fields, but incorrectly concluded that the interaction between cavities is repulsive.
The interaction between two pressurized, equal cavities in an isotropic, linear elastic medium is, in fact, always attractive, irrespective of their separation (as will be verified later in this work).

Although the linear elastic problem is clearly well studied, extending these findings to account for the complexities introduced by nonlinear elasticity, which is characteristic of soft materials such as elastomers, polymers, and biological tissues, remains an open challenge.
In contrast to linear elastic materials, these materials exhibit nonlinear stress-strain relationships that lead to rich and often counterintuitive mechanical behaviors.

We aim to address this challenge by investigating the elastic interaction of two pressurized cylindrical cavities embedded in a two-dimensional hyperelastic medium. 
We focus on three widely-used material models --- neo-Hookean, Mooney-Rivlin, and Arruda-Boyce --- to elucidate the role of constitutive nonlinearity and strain-stiffening in the interaction between pressurized cylindrical cavities \citep{rivlin1948large, mooney1940theory, Arruda1993}.
Our approach differs from prior studies by framing the interaction in terms of driving forces derived from total potential energy, enabling precise quantification of attraction and repulsion mechanisms.
Our results reveal that nonlinear elasticity introduces a critical separation distance where the interaction transitions from attraction to repulsion --- a behavior absent in linear elasticity. Additionally, we demonstrate that strain-stiffening parameters, particularly in the Arruda-Boyce model, significantly influence the interaction, offering a potential pathway for controlling cavity configurations through material and loading parameters.

The two-dimensional framework examined here is relevant to several cutting-edge applications, including phase separation in confined systems \citep{seol2003computer, binder2010phase}, membrane-mediated interactions between cylindrical inclusions in lipid bilayers \citep{Bohinc2003}, and the design of parallel tunnels in geomechanics \citep{kim2004interaction, fu2015analytical, zeng2023analytical}.  
Moreover, these insights could qualitatively inform analogous three-dimensional problems involving spherical cavities, further broadening the scope of this research.    

\section{Problem Setup and Analysis}
\subsection{Geometry, Loading, Boundary Conditions, and Kinematics}
\begin{figure}[ht]
\centering

  \includegraphics[scale = 0.25]{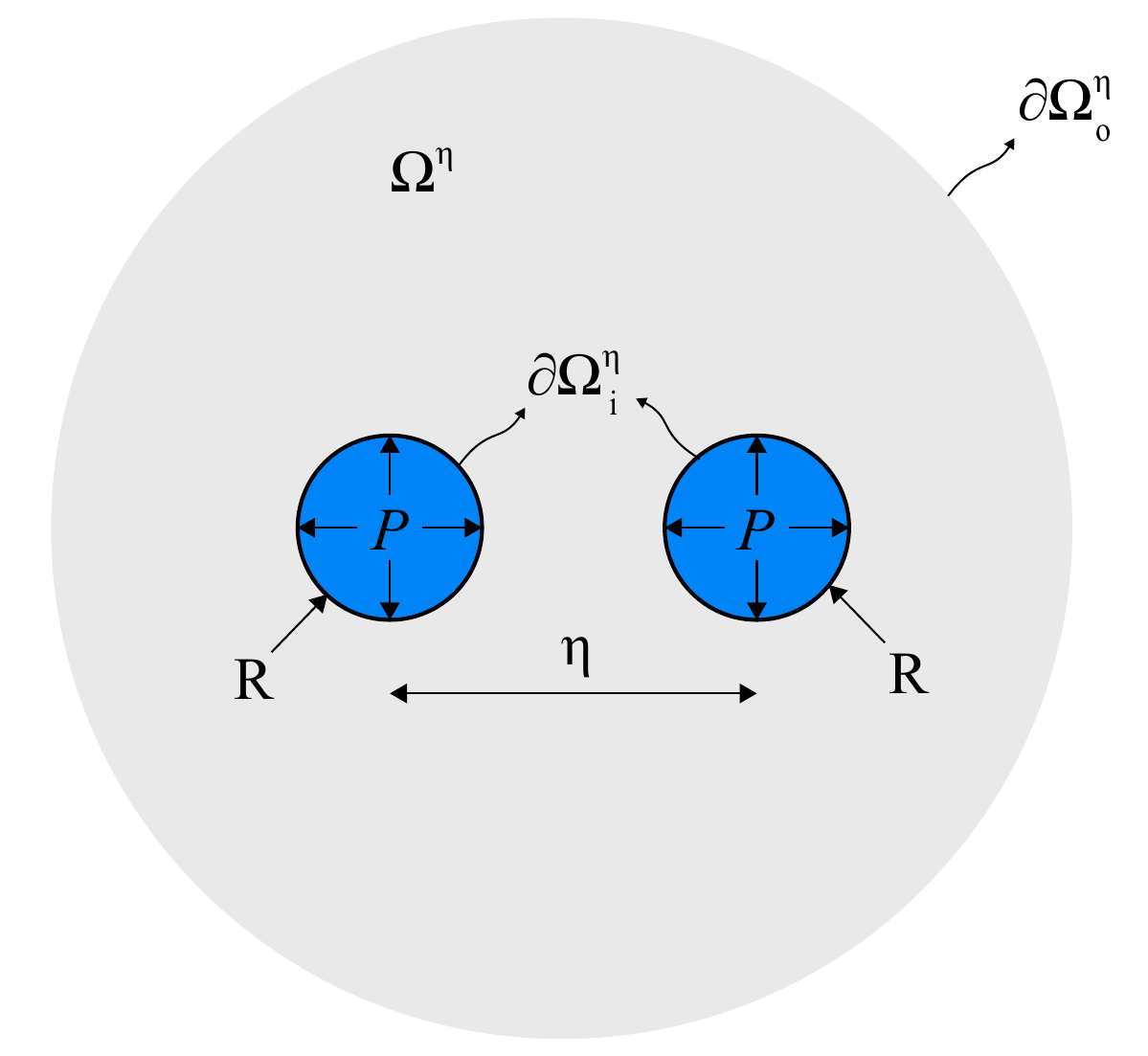}\\
  \caption{Schematic of the two-cavity system in a hyperelastic medium under constant internal pressure  P.}
  \label{Geometry}
\end{figure}

\noindent Figure \ref{Geometry} shows the geometry of the problem with two circular cavities inside an infinite, elastic matrix.
The internal boundaries of the domain, which correspond to the two cavities, are collectively denoted by $\partial \Omega^{\eta}_{i}$.
The outer boundary of the infinite domain is denoted by $\partial \Omega^{\eta}_{o}$.
A plane-strain setting is considered, which simplifies three-dimensional effects and provides tractability, while capturing key two-dimensional physics relevant to soft materials and structural systems.

Both cavities have the same undeformed radius $R$.
The internal pressure within the cavities in the  deformed configuration, $P$, is assumed to be constant throughout the deformation process.
Such a loading could represent scenarios where the pressure is explicitly applied or maintained through a process like phase separation\footnote{A constant pressure setting, however, would not be applicable to problems where the pressure changes with expansion, such as expansion of a cavity containing gas, where the pressure and volume are coupled together by gas law.}.
The outer boundary of the domain (at infinity) is taken to be traction-free.
The traction boundary conditions can be cast in the following form:
\begin{equation}
\begin{cases} 
\displaystyle \bfsigma \bfn = -P\bfn \text{ for } \bfx \in \partial \Omega^{\eta}_{i},\\
\displaystyle \bfsigma \bfn = \mathbf{0} \text{ for } \bfx \in \partial \Omega^{\eta}_{o}.
\end{cases}
\end{equation}
where the material points in the body in the reference configuration are denoted by $\X$, and the corresponding points in the deformed configuration are denoted by $\bfx$, which gives the displacement field  as $\bfu =\bfx -\bfX$.

The deformation gradient, $\bfF = \pards{\bfx}{\bfX}$, is a map from the reference configuration to the deformed configuration.
Finally, we introduce the Left Cauchy-Green deformation tensor, \bfB, and its two invariants $\mathbb{I}_1, \mathbb{I}_2$ as,
\begin{align}
	\bfB =& \bfF\bfF^\top\\
	\mathbb{I}_1 =& \operatorname{tr}(\bfB)\\
	\mathbb{I}_2 =&\frac{1}{2}\left((\operatorname{tr}(\mathbf{B}))^2 - \operatorname{tr}(\mathbf{B}^2)\right)
\end{align}
for the purpose of defining constitutive models in section \ref{sec:2.2}.

\subsection{Nonlinear Elastic Material Models}\label{sec:2.2}
\noindent We choose three commonly used isotropic, hyperelastic, incompressible material models to analyze in this study: neo-Hookean (NH), Mooney-Rivlin (MR), and Arruda-Boyce (AB) models. Their strain energy density functions are defined as below.

\begin{gather}
    \psi_{\text{NH}} = \frac{\mu}{2}(\mathbb{I}_1 - 3) \quad \label{eq:W_NH}\\
\psi_{\text{MR}} = \alpha\frac{\mu}{2} (\mathbb{I}_1 - 3) + \left(1-\alpha\right)\frac{\mu}{2} (	\mathbb{I}_2 - 3) \label{eq:W_MR}\\
\psi_{\text{AB}} = C_{2} \Big[\frac{1}{2}(	\mathbb{I}_1 - 3) + \frac{1}{20 \lambda_m^2} (\mathbb{I}_1^2 - 9) + \frac{11}{1050 \lambda_m^4} (\mathbb{I}_1^3 - 27) + \frac{19}{7000 \lambda_m^6} (\mathbb{I}_1^4 - 81) + \frac{519}{673750 \lambda_m^8} (	\mathbb{I}_1^5 - 243)\Big] 
    \label{eq:W_AB}
\end{gather}

where $\mu$ is the shear modulus, and the parameter $0 < \alpha < 1$ controls the degree of strain stiffening in Mooney-Rivlin model \cite{Kanner2008}. 
The consistency condition \cite{Arruda1993} for the incompressible Arruda-Boyce model is

\begin{gather}
    \mu = C_2 \left( 1 + \frac{3}{5 \lambda_m^2} + \frac{99}{175 \lambda_m^4} + \frac{513}{875 \lambda_m^6} + \frac{42039}{67375 \lambda_m^8} \right).
\end{gather}
Among the three models, Mooney-Rivlin and Arruda-Boyce models show strain stiffening, while neo-Hookean does not.

\subsection{Driving Force}
\noindent The total potential energy of the system (per unit length for plane strain) can be written as,
\begin{equation}
	\Pi^{\eta}[\bfu] = \int_{\Omega^{\eta}}\psi \mathrm{\ d\Omega} - 2P \Delta A, \label{PE}
\end{equation} 
where $\bfu$ is the displacement field, $\psi \equiv \psi(\bfF)$ is the strain energy density in the deformed configuration, $P$ is the pressure on the inner boundary $\partial \Omega^{\eta}_{i}$ in the deformed configuration of the domain, and $\Delta A$ is the change in area of the cavity due to deformation.
For every admissible $\eta$, the equilibrium displacement field and the associated potential energy are denoted by $\bfu^{\eta}_{eqm}$ and $\Pi^{\eta}_{eqm} := \Pi^{\eta}[\bfu^{\eta}_{eqm}]$, respectively, with both $\bfu^{\eta}_{eqm}$ and $\Pi^{\eta}_{eqm}$ being implicitly dependent on $\eta$.

The driving force $\mathcal{F}$ can then be written as,
\begin{equation}
	\mathcal{F} = -\frac{d\Pi^{\eta}_{eqm}}{d\eta} \label{driving_force}
\end{equation}
This driving force quantifies both the magnitude and nature of the interaction between the cavities, indicating whether it is attractive or repulsive.


\subsection{Computational Approach}
\noindent We simulate this problem using a commercial finite element software ABAQUS \cite{Abaqus}.
The geometry with domain radius 5 and cavity radius 0.1 is discretized using bilinear quadrilateral plane strain reduced hybrid elements (CPE4RH).
For each material model, a sequence of meshes with $\eta/R \in [2.2, 20]$ is constructed in steps of $\eta/R = 0.2$.
Each of these models is simulated for a range of pressures $P/\mu \in [0.25, 1.75]$ in steps of $P/\mu = 0.25$.
From each simulation run, we obtain all displacements, stresses, deformation gradient, and total strain energy over the entire domain.
This data is post-processed to obtain potential energy using equation \eqref{PE}, including the deformed area.
A smooth curve is then fitted to potential energy data to compute its derivative for calculating the driving force.
Mesh convergence studies are reported in Appendix C to ensure that the chosen mesh was appropriately fine.
A representative deformed area trend is reported in Appendix D.
The complete, non-fitted potential energy data is reported in Appendix E.

\begin{figure}[ht]
\centering
    \includegraphics[scale =0.25] {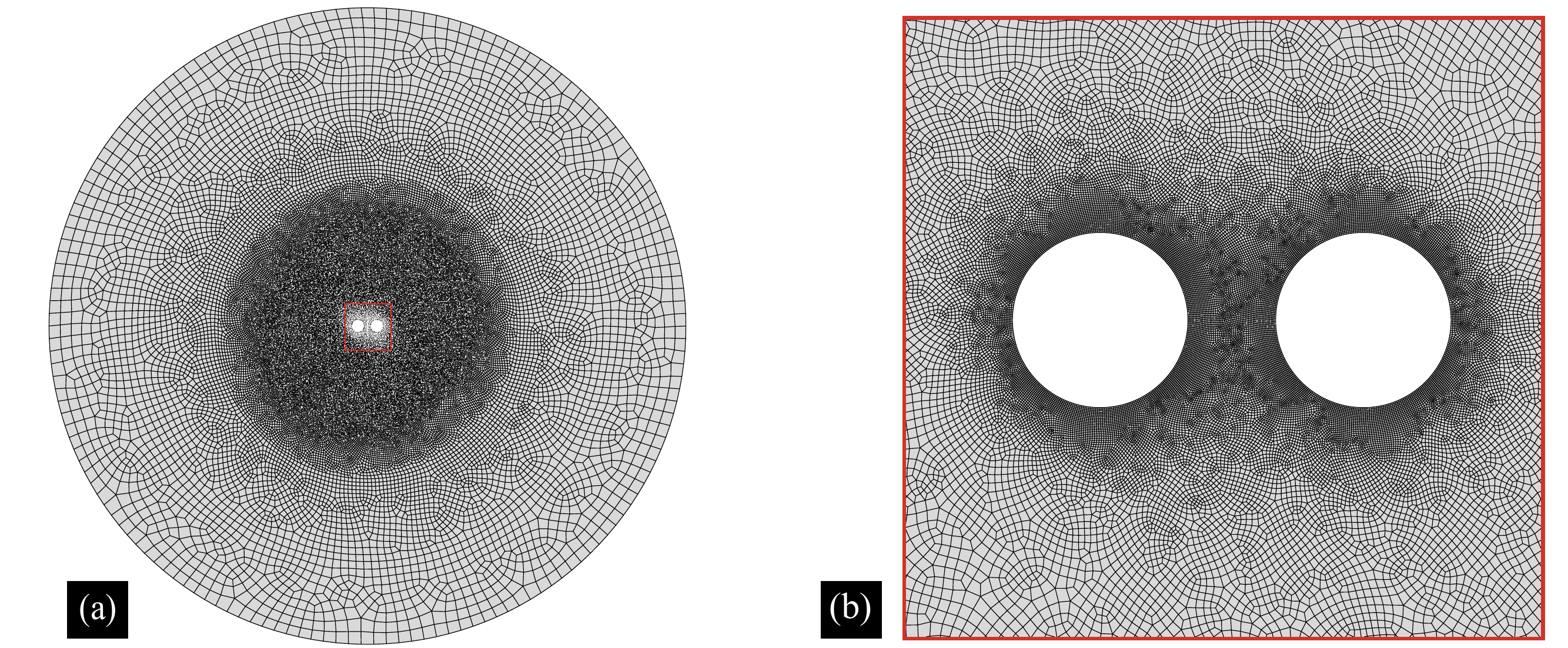}
\caption{A representative generated mesh: (a) entire domain and (b) locally refined region for $\eta/R=3$ using 75182 CPE4RH elements.}
\label{dicretized_geometry}
\end{figure}

\section{Results and Discussion}

\subsection{Linear Elastic Medium}
\begin{figure}[ht]
\centering
	 \includegraphics[scale = 0.25]{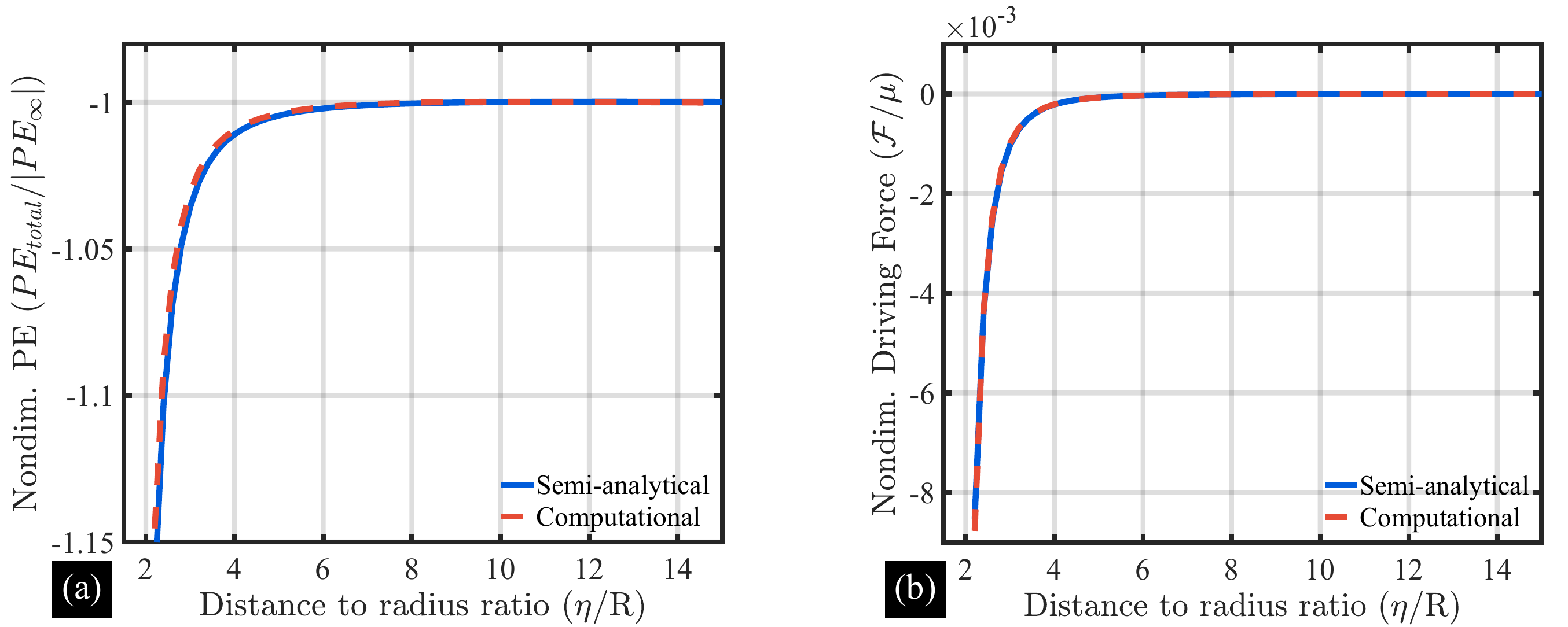}
	\caption{\small (a) Potential energy and (b) driving force for two pressurized cavities in a linear elastic medium. The results are independent of $P/\mu$.}
    \label{PE_DF_LE}
\end{figure}

\noindent We start by validating the proposed computational methodology by applying it to the case of two pressurized cavities inside a linear elastic medium.
A full series solution is provided in the Appendix A, obtained using the bipolar coordinate system\footnote{As noted in the Introduction, we have found errors in the solutions presented in \cite{Ling1948,Davanas1992} but their overall approach is correct. As a resource for readers, we include the corrected solution and its MATLAB implementation in Appendix A.}.
We find that the strain energy and driving force computed from FEM are in excellent agreement with the semi-analytical (series) solution, as shown in Figure \ref{PE_DF_LE}.
The potential energy is non-dimensionalized by the $PE_{\infty}$, which is the potential energy of the system when the two cavities are infinitely far away from each other (see Appendix A).
The driving force on the cavities is always negative, which means that the cavities are always \textit{attracted} to each other, irrespective of the distance between them.

\subsection{Nonlinear Elastic Medium}
\begin{figure}[htb!]
\centering
      \includegraphics[scale = 0.24]{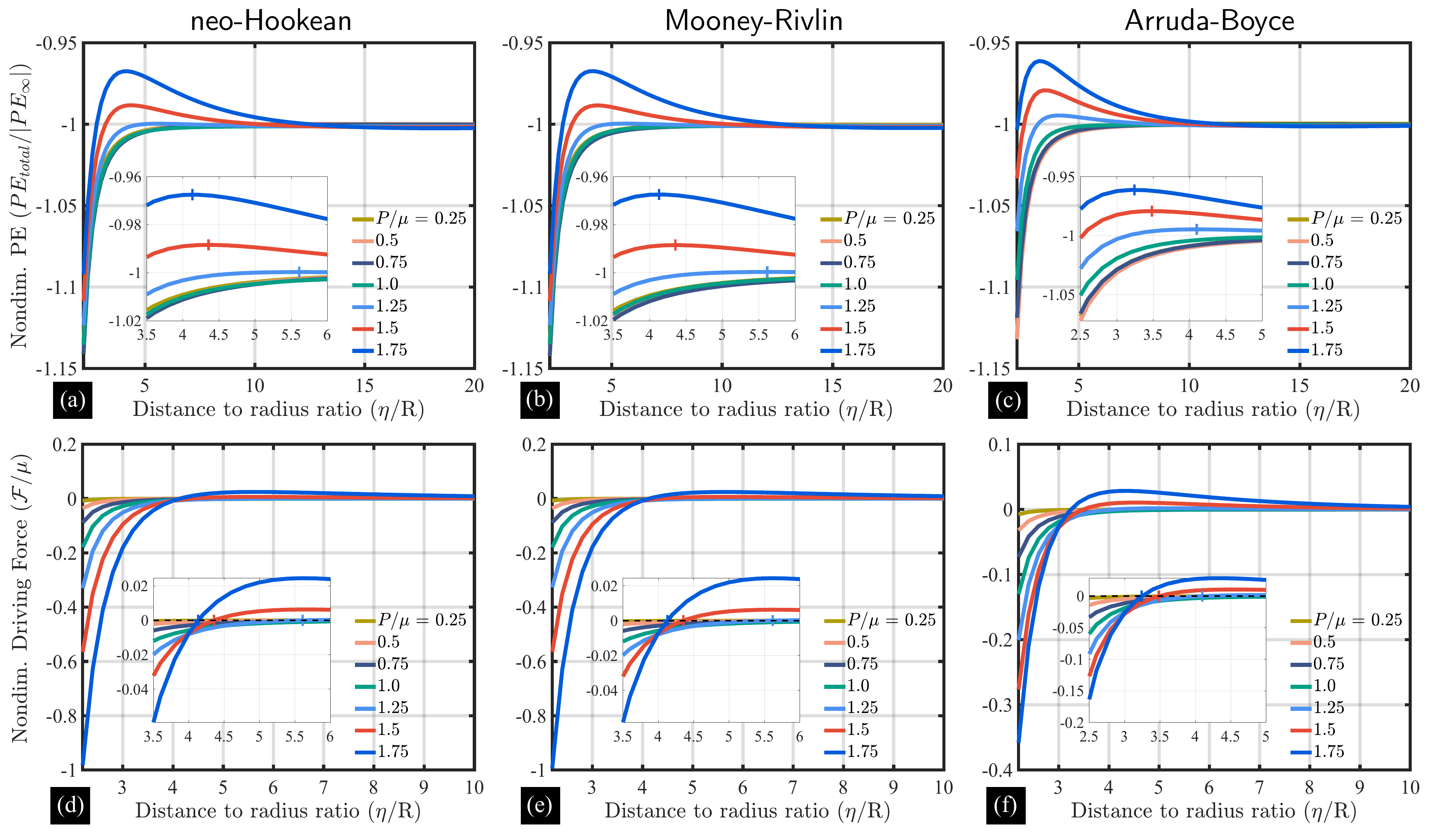}
 \caption{Summary of potential energy and driving force for all the models ($\alpha=0.5$, $\lambda_m=2$).}    \label{PE_DF_combined}
\end{figure}

\noindent To compare and contrast the effects of nonlinear elasticity and strain-stiffening on the interaction between the cavities, the results are presented in a non-dimensionalized form and the three materials models (NH, MR and AB) are taken to have the same shear modulus.
Figure \ref{PE_DF_combined} shows the potential energy and driving force for different material models for a range of pressures.

\begin{figure}[htb!]
\centering
       \includegraphics[scale = 0.24] {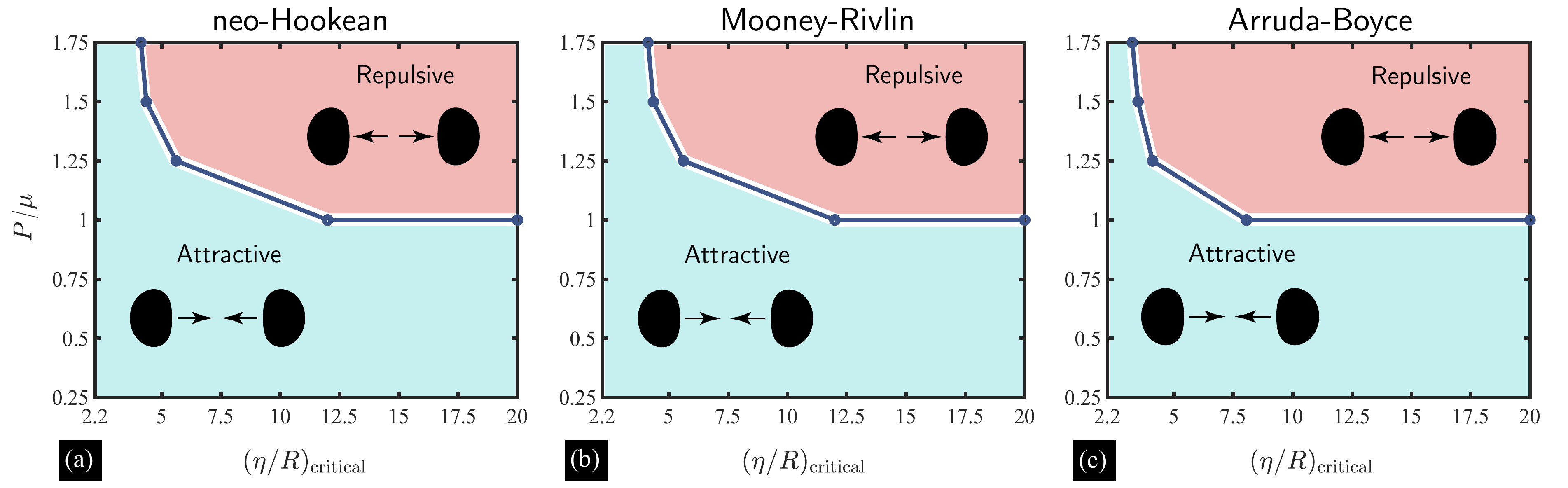}
\caption{Phase diagram of nonlinear elastic material models ($\alpha=0.5$, $\lambda_m=2$).} 
\label{Phase_diagram} 
\end{figure}

Across all models, potential energy and driving force is monotonic for low pressures ($P/\mu \lesssim 1$), revealing that the cavities are always attracted to each other at low pressures; the strength of attraction decreases with increasing separation between the cavities.
This behavior is consistent with the linear elastic trend.

However, at higher pressures ($P/\mu \gtrsim 1$), nonlinear elasticity changes the energy landscape significantly.
In stark contrast to linear elastic case, we find that the potential energy reaches a maximum at a critical separation between the cavities, $(\eta/R)_{critical}$, and thus, the driving force changes signs and becomes positive for $\eta/R > (\eta/R)_{critical}$.
If the cavities are further apart than the critical separation, they \textit{repel} each other and if they are closer, they attract.
At the critical separation, driving force is zero and represents an unstable configurational equilibrium for the non-linear elastic system at higher pressures.
$(\eta/R)_{critical}$ varies with pressures, as shown in figure \ref{Phase_diagram}, for all three models.

\begin{figure}[h]
\centering
    \includegraphics[scale = 0.25] {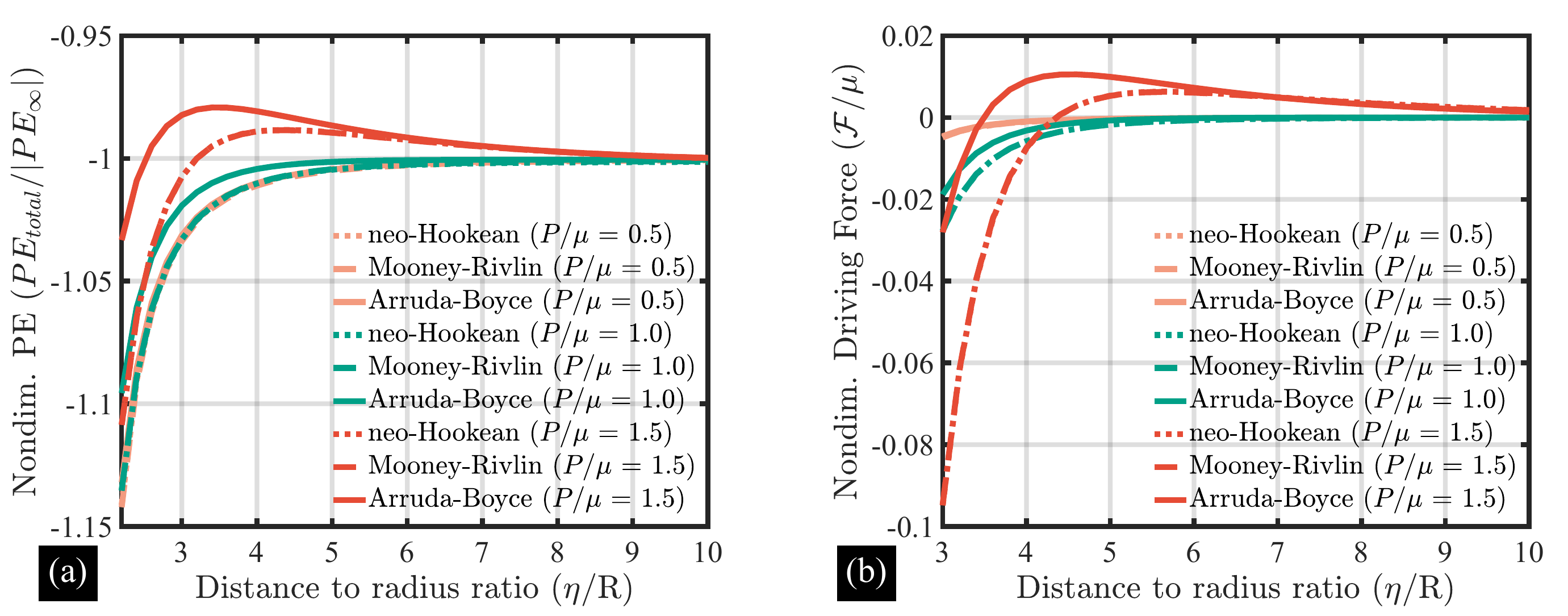}
\caption{Comparison of (a) potential energy and (b) driving force for nonlinear elastic material models ($\alpha=0.5$, $\lambda_m=2$).}
\label{PE_DF_comp}
\end{figure}%

Figure \ref{PE_DF_comp} shows the comparison between the different constitutive models for representative values of $\alpha =0.5$ and $\lambda_m = 2$.
neo-Hookean and Mooney-Rivlin models do not show any significant differences, whereas the Arruda-Boyce model shows noticeable differences from the other two models.

\subsection{Effect of Strain Stiffening}
\noindent Many soft materials, particularly biological materials like arteries, skin, and actin filaments in cytoskeleton, exhibit increasing stiffness with deformation \citep{horgan2006phenomenological}.
This behavior  --- better known as strain-stiffening --- is believed to hold physiological importance in preventing damage to biological materials \citep{storm2005nonlinear,erk2010strain}.
\begin{figure}[htb!]
\centering
       \includegraphics[scale = 0.25] {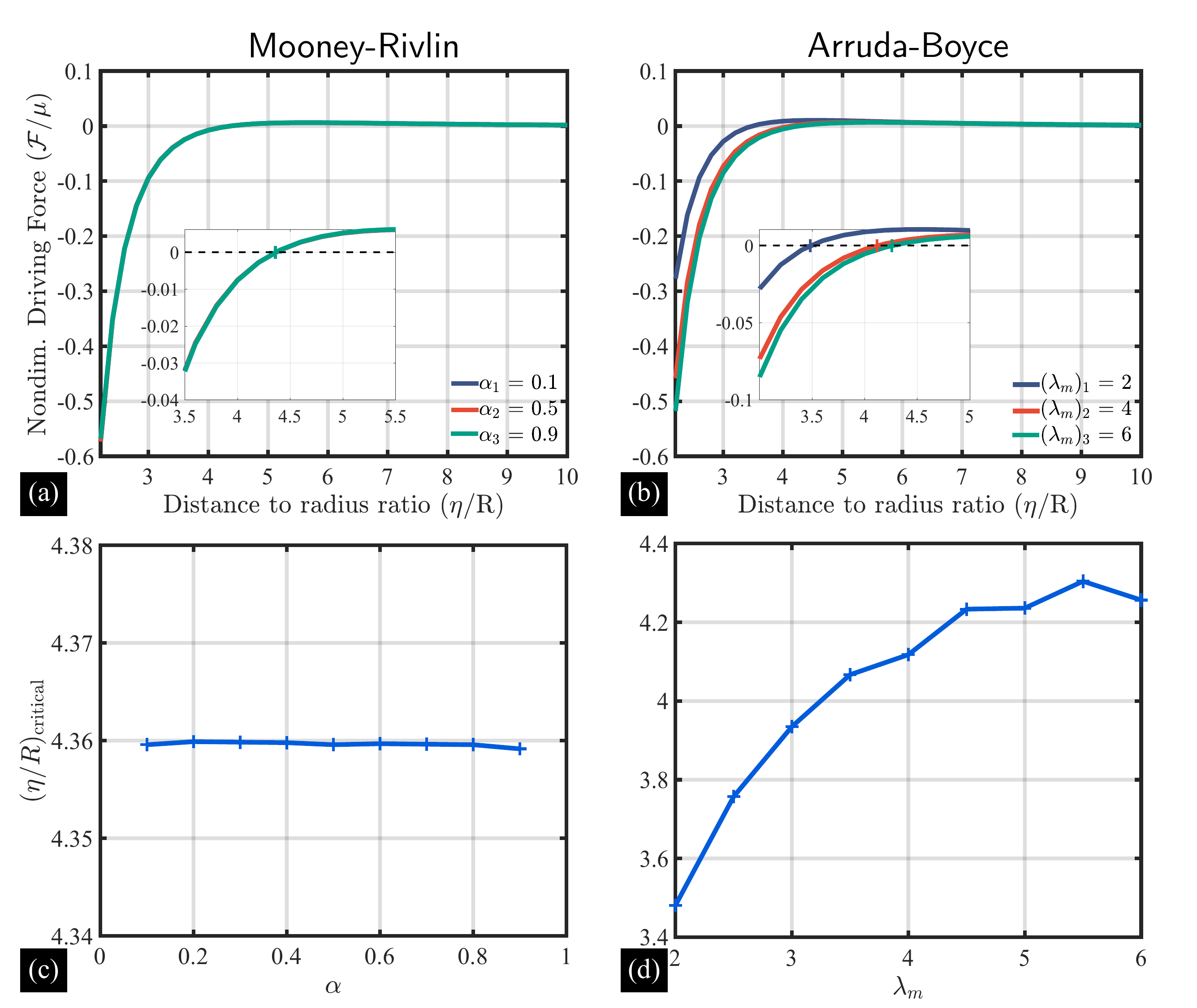}
\caption{Driving force variation for different strain-stiffening parameters ($P/\mu=1.5$).} 
\label{DF_var} 
\end{figure}
Among our chosen models, Mooney-Rivlin and Arruda-Boyce models can capture strain-stiffening, which can be controlled by the parameters $\alpha$ and $\lambda_m$, respectively.
Figure \ref{DF_var} shows the effect of strain-stiffening parameters by comparing the driving forces for a representative case of $P/\mu = 1.5$.

Interestingly, Mooney-Rivlin model shows nearly no dependence on strain stiffening for the entire range of $\alpha$, while Arruda-Boyce model does.
In the latter, a smaller limiting stretch ($\lambda_m$) leads to larger driving forces and smaller $(\eta/R)_{critical}$ values.
The differences between the two models can attributed to the specific state of stress that arises in the two-cavity problem --- such a loading results in regions of severe compression and tension and large shear stresses, particularly around and between the cavities, and Arruda-Boyce model shows more prominent strain-stiffening behavior for these loadings as compared to the Mooney-Rivlin model \citep{boyce2000constitutive, suchocki2021polyconvex}.

\section{Conclusion}
\noindent Through extensive semi-analytical and finite element analysis, we find that the interaction between two pressurized, equal cavities in an infinite elastic medium is always attractive at low pressures ($P/\mu \lesssim 1.0$), even for the non-linear elastic material models tested in this study (neo-Hookean, Mooney-Rivlin, and Arruda-Boyce).
Non-linear elasticity affects the interaction at higher pressures ($P/\mu \gtrsim 1.0$), where the potential energy becomes non-monotonic with inter-cavity separation. 
For inter-cavity separation smaller than the critical separation, the driving force is negative, implying attractive interaction between the cavities; for inter-cavity separation larger than the critical separation, driving force is positive, implying that the cavities repel each other.
The energy landscape of the Mooney-Rivlin model appears to be almost independent of strain-stiffening, and its quantitative trends remain nearly the same as the neo-Hookean model, whereas the Arruda-Boyce model exhibits a more prominent dependence on strain-stiffening. 
A higher strain stiffening in Arruda-Boyce model leads to smaller critical inter-cavity separation.
These findings provide valuable insights into the mechanics of pressurized cavities in elastic media, with implications for engineering applications involving poroelasticity, phase separation, cavitation, and the design of soft materials. 
Future work could extend this analysis to consider anisotropic materials, time-dependent effects, and multi-cavity interactions, broadening the applicability of the results to more complex systems.

\section{Acknowledgement}
\noindent 
Support from NH BioMade under the National Science Foundation
EPSCoR award \#1757371 is acknowledged.
Insightful discussions with Prof. Haneesh Kesari (Brown University) are acknowledged.
\newpage
\section*{APPENDIX}
\appendix
\section{Linear Elastic Solution}
\begin{figure}[ht]
\centering
  \includegraphics[scale = 0.25]{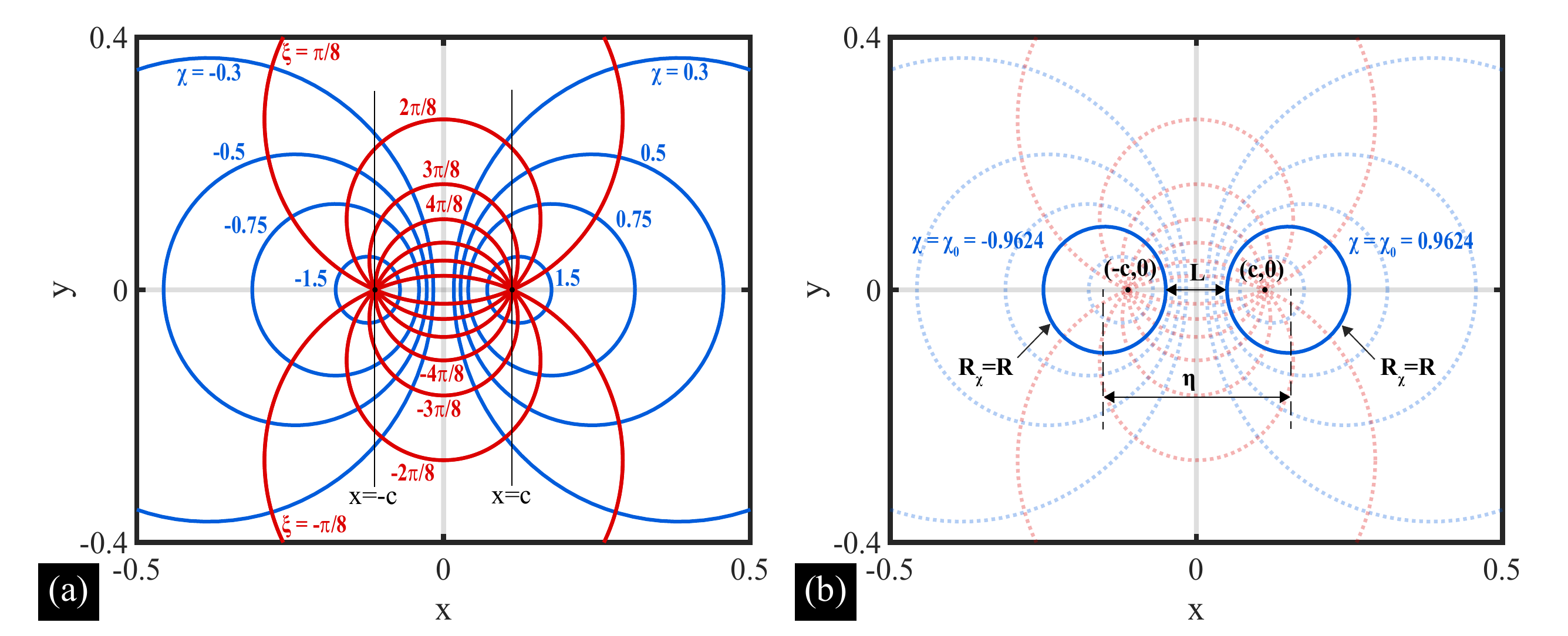}
  \caption{Bipolar coordinate system and the geometry used in linear elastic problem $(R=0.1, \eta=3R, L=R).$}
  \label{Bipolar}
\end{figure}

Closely following the formulation presented by Ling \cite{Ling1948} (and followed by Davanas \cite{Davanas1992}), an analytical solution has been derived using a bipolar coordinate system. A point in Cartesian coordinate (x,y) is mapped to a bipolar coordinate $(\chi, \xi)$ using the below equations  \cite{Lucht2015}

\begin{equation}
\begin{gathered}
    x = J \sinh{\left(\chi\right)} \\
    y = J sin{\left(\xi\right)}
\end{gathered}
\label{eq:xy}
\end{equation}

where $J$ is the Jacobian of this transformation given by
{\footnotesize
\begin{equation}
    J = \frac{c}{\cosh{\left(\chi\right)}-\cos{\left(\xi\right)}}
\end{equation}
}
A schematic of the bipolar system is shown in figure \ref{Bipolar}. 
An infinite number of circles, blue and red, respectively in the (x,y) plane, defined by equation \ref{eq:xy}, are given by

\begin{equation}
\begin{gathered}
    (x - \bar{x}) + y^2 = R_{\chi}^2 \\
    x^2 + (y - \bar{y})^2 = R_{\xi}^2
\end{gathered}
\end{equation}
where $\bar{x}$, $\bar{y}$ are center points of the circle, $R_{\chi}$ and $R_{\xi}$ are circle radius defined by
{\footnotesize
\begin{equation}
\begin{gathered}
    x_c = a \coth(\chi) \\
    y_c = a \cot(\xi) \\
    R_{\chi} = \frac{c}{\sinh(\chi)} \\
    R_{\xi} = \frac{c}{\sin(\xi)}
\end{gathered}
\end{equation}
}
with below relationship between focal points $(-c,0)$ and $(c,0)$, separation distance $L$, and radius $R_{\chi}$
{\footnotesize
\begin{gather}
    c^2 + R_{\chi}^2 = \left(\frac{L}{2} + R_{\chi}\right)^2
\end{gather}
}
The geometry used to describe the current problem is highlighted in figure \ref{Bipolar}.

In this problem, total potential energy is calculated using equation \eqref{PE}.
To calculate strain energy, stress components must be determined. 
Stress components in a bipolar coordinate system are calculated as follows,

{\footnotesize
\begin{align}
\begin{split}
    \sigma_{\chi\chi} = & \Bigg\{-\frac{1}{2}K[\cosh{(2\chi)}+\cos{(2\xi)}-2\cosh{(\chi)}\cos{(\xi)}] + \sum_{n=1}^{N}A_n\Big[-n^2\cosh{((n+1)\chi)}[\cosh{(\chi)}-\cos{(\xi)}]\cos{(n\xi)} \\
    & -(n+1)\sinh{((n+1)\chi)}\sinh{(\chi)}\cos{(n\xi)}+n\sin{(\xi)}\sin{(n\xi)}\cosh{((n+1)\chi)} +\cos{(n\xi)}\cosh{(\chi)}\cosh{((n+1)\chi)}\Big] \\
    & + B_1 + \sum_{n=2}^{N}B_n\Big[-n^2\cosh{((n-1)\chi)}[\cosh{(\chi)}-\cos{(\xi)}]\cos{(n\xi)} -(n-1)\sinh{((n-1)\chi)}\sinh{(\chi)}\cos{(n\xi)} \\
    & + n\sin{(\xi)}\sin{(n\xi)}\cosh{((n-1)\chi)} +\cos{(n\xi)}\cosh{(\chi)}\cosh{((n-1)\chi)}\Big]\Bigg\}c^{-1}
\end{split}
\end{align}
}

{\footnotesize
\begin{align}
\begin{split}
    \sigma_{\xi\xi}= & \Bigg\{\frac{1}{2}K[\cosh{(2\chi)}+\cos{(2\xi)}-2\cosh{(\chi)}\cos{(\xi)}]\ +\sum_{n=1}^{N}A_n\Big[(n+1)^2\cosh{((n+1)\chi)}[\cosh{(\chi)}-\cos{(\xi)}]\cos{(n\xi)}  \\
    & -(n+1)\sinh{(\left(n+1\right)\chi)}\sinh{(\chi)}\cos{(n\xi)}+n\sin{(\xi)}\sin{(n\xi)}\cosh{((n+1)\chi)} + \cos{(n\xi)}\cos{(\xi)}\cosh{((n+1)\chi)}\Big] \\
    & + B_1 +\sum_{n=2}^{N}B_n\Big[(n-1)^2\cosh{((n-1)\chi)}[\cosh{(\chi)}-\cos{(\xi)}]\cos{(n\xi)} -(n-1)\sinh{((n-1)\chi)}\sinh{(\chi)}\cos{(n\xi)} \\
    & + n\sin{(\xi)}\sin{(n\xi)}\cosh{((n-1)\chi)} + \cos{(n\xi)}\cos{(\xi)}\cosh{((n-1)\chi)}\Big]\Bigg\} c^{-1}
\end{split}
\end{align}
}

{\footnotesize
\begin{align}
\begin{split}
    \sigma_{\chi\xi}= & -K\sinh{(\chi)}\sin{(\xi)}
     +\sum_{n=1}^{N}{A_n\Big[n(n+1)\sinh{((n+1)\chi)}[\cosh{(\chi)}-\cos{(\xi)}]\sin{(n\xi)}\Big]} \\
    & +\sum_{n=2}^{N}{B_n\Big[n(n-1)\sinh{((n-1)\chi)}[\cosh{(\chi)}-\cos{(\xi)}]\sin{(n\xi)}\Big]}
\end{split}
\end{align}
}
where $K$, $A_n$, $B_1$ and $B_n$ are defined as
{\footnotesize
\begin{gather}
    K = \frac{cP}{\frac{1}{2}+\tanh{\left(\chi_0\right)}\sinh^2{\left(\chi_0\right)}-4F} \\
    A_n = \frac{2K\left[e^{-n\chi_0}\sinh{\left(n\chi_0\right)}+ne^{-\chi_0}\sinh{\left(\chi_0\right)}\right]}{n\left(n+1\right)\left[\sinh{\left(2n\chi_0\right)}+n\sinh{\left(2\chi_0\right)}\right]} \\
    B_1 = \frac{1}{2}\left[K\tanh{\left(\chi_0\right)}\cosh{\left(2\chi_0\right)}-2cP\right] \\   
    B_n = \frac{-2K\left[e^{-n\chi_0}\sinh{\left(n\chi_0\right)}+ne^{\chi_0}\sinh{\left(\chi_0\right)}\right]}{n\left(n-1\right)\left[\sinh{\left(2n\chi_0\right)}+n\sinh{\left(2\chi_0\right)}\right]}
\end{gather}
}
and $c$, $\chi_0$ and $F$ are calculated as

{\footnotesize
\begin{gather}
    c = \sqrt{\left(\frac{L}{2}+R\right)^2-R^2} \\
    \chi_0 = {\sinh}^{-1}{\left(\frac{c}{R}\right)} \\
    F = \sum_{n=2}^{N}\frac{e^{-n\chi_0}\sinh{\left(n\chi_0\right)}+n\sinh{\left(\chi_0\right)}\left[n\sinh{\left(\chi_0\right)+\cosh{\left(\chi_0\right)}}\right]}{n\left(n^2-1\right)\left[\sinh{\left(2n\chi_0\right)+n\sinh{\left(2\chi_0\right)}}\right]}
\end{gather}
}
More details can be found in \cite{Ling1948} and \cite{Davanas1992}. 
The total strain energy is evaluated once the stress components have been determined.
{\footnotesize
\begin{align}
\begin{split}
    SE_{tot}= & 2\left(\int_{y=-\infty}^{\infty}\int_{x=0}^{\infty}{\frac{1}{2}\sigma_{ij}\varepsilon_{ij}\ dx\ dy}\right)
    =\int_{\chi=0}^{\chi_0}\int_{\xi=-\pi}^{\pi}{\sigma_{ij}\varepsilon_{ij}\ J^2d\xi\ d\chi} \\
    & =\sum_{\chi=0}^{\chi_0}\sum_{\xi=-\pi}^{\pi}{\left(\sigma_{\chi\chi}\varepsilon_{\chi\chi}+\sigma_{\xi\xi}\varepsilon_{\xi\xi}+2\sigma_{\chi\xi}\varepsilon_{\chi\xi}\right)J^2}    
\end{split}
\end{align}
}
The stress-strain relationship in linear elasticity for plane strain condition is as follows,
{\footnotesize
\begin{align}
\sigma_{11} &= \frac{E}{(1+\nu)(1-2\nu)} \left[ (1-\nu) \epsilon_{11} + \nu \epsilon_{22} \right], \\
\sigma_{22} &= \frac{E}{(1+\nu)(1-2\nu)} \left[ \nu \epsilon_{11} + (1-\nu) \epsilon_{22} \right], \\
\sigma_{12} &= \mu \gamma_{12},
\end{align}
}
and in the inverted form, strain-stress is written as:
{\footnotesize
\begin{align}
\epsilon_{11} &= \frac{1}{E} \left[ \sigma_{11} - \nu (\sigma_{22} + \nu (\sigma_{11} + \sigma_{22})) \right] = \frac{1}{E} \left[ \sigma_{11} (1 - \nu^2) - \nu \sigma_{22} (1 + \nu) \right], \\
\epsilon_{22} &= \frac{1}{E} \left[ \sigma_{22} - \nu (\sigma_{11} + \nu (\sigma_{11} + \sigma_{22})) \right] = \frac{1}{E} \left[ \sigma_{22} (1 - \nu^2) - \nu \sigma_{11} (1 + \nu) \right], \\
\gamma_{12} &= \frac{\tau_{12}}{\mu} = \frac{2(1 + \nu)}{E} \tau_{12},
\end{align}
}
Incorporating plane-strain conditions and rewriting the equation in terms of stress components, the final equation for total strain energy becomes

{\footnotesize
\begin{align}
    SE_{tot} = & \sum_{\chi=0}^{\chi_0}\sum_{\xi=-\pi}^{\pi}\Bigg\{\sigma_{\chi\chi}\left[\frac{1-\nu^2}{E}\sigma_{\chi\chi}-\frac{\nu(1+\nu)}{E}\sigma_{\xi\xi}\right] \nonumber
    +\sigma_{\xi\xi}\left[-\frac{\nu(1+\nu)}{E}\sigma_{\chi\chi}+\frac{1-\nu^2}{E}\sigma_{\xi\xi}\right] \nonumber \\
    & +2\sigma_{\chi\xi}\left[\frac{(1+\nu)}{E}\right]\Bigg\}\left(\frac{c}{\cosh{\left(\chi\right)}-\cos{\left(\xi\right)}}\right)^2
\end{align}
}
In the current problem (linear elastic with static equilibrium), potential energy of external forces can be computed as $-2SE_{tot}$, thus giving the total potential energy as
{\footnotesize
\begin{equation}
    PE_{tot} = -SE_{tot}
\end{equation}
}
The total potential energy of a single pressurized cavity ($PE_0$) is equal to half of the energy of the system of two cavities at infinite separation ($PE_{\infty}$), which is calculated in 2D as \cite{Davanas1992}
{\footnotesize
\begin{equation}
    PE_{\infty} = 2PE_0 = 2\frac{\pi P^2R^2}{2\mu}
\end{equation}
}
So, \textit{semi-analytical} total potential energy is non-dimensionalized as
{\footnotesize
\begin{equation}
    PE_{norm} = \frac{PE_{tot}}{|PE_{\infty}|}
\label{eq:PE_norm}
\end{equation}
}
On the other hand, \textit{computational }total potential energy is calculated by obtaining the total strain energy and the area change from finite element simulation and plugging into equation \eqref{PE}. 
It is also non-dimensionalized by dividing by twice the potential energy of a single pressurized cavity ($2PE_0$).


\section{Pressurized single cavity: validation of analytical and computational approach}
\label{appendix_A}
\begin{figure}[ht]
\centering
  \includegraphics[scale=0.6]{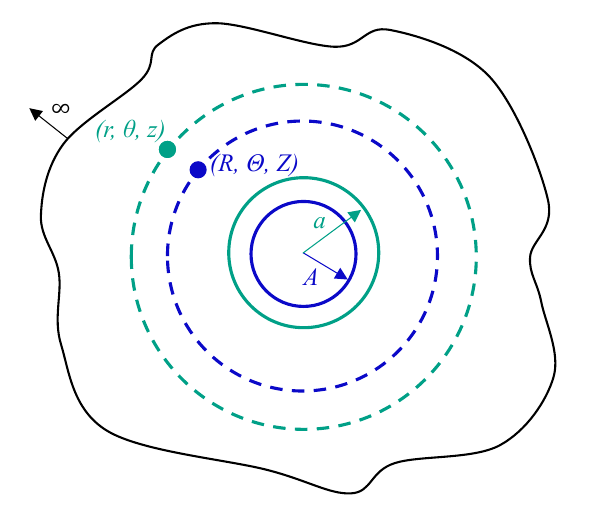}\\
  \caption{Single cavity with reference (blue) and deformed (green) coordinates.}
  \label{CylCoord}
\end{figure}

\noindent The geometry of the 2D problem in a cylindrical coordinate system is illustrated in figure \ref{CylCoord}.
Body points in original configuration are denoted as (\textit{R},$\mathrm{\Theta}$,\textit{Z}) which are transformed to deformed configuration $\left(r,\theta,z\right)$. 
The undeformed radius of the circular cavity is $A$, and the deformed radius is $a$.
Due to symmetry $\mathrm{\Theta}=\theta$ and as the domain is two-dimensional \textit{Z}=\textit{z}, results in a deformation from $\left(R,\theta,z\right)$ to $\left(r,\theta,z\right)$ when only radial components are changing.
Due to the incompressibility condition, the following conclusion can be derived (\textit{V} and \textit{v} are initial and final volumes),

\begin{equation}
    \begin{aligned}
    &\textit{J} = \textit{det}\, \mathbf{F} = \frac{dv}{dV} = 1 \\
    &\pi \left( R^2 - A^2 \right) = \pi \left( r^2 - a^2 \right) \quad \\
    &\quad r = \sqrt{R^2 + a^2 - A^2} \quad \\
    &\quad \frac{\partial r}{\partial R} = \frac{R}{r}
    \end{aligned}
\end{equation}

The deformation gradient in a cylindrical coordinate system is calculated as,
{\footnotesize
\begin{equation}
    \mathbf{F} = 
    \begin{bmatrix}
    \frac{\partial r}{\partial R} & \frac{1}{R} \frac{\partial r}{\partial \Theta} & \frac{\partial r}{\partial Z} \\
    r \frac{\partial \theta}{\partial R} & \frac{r}{R} \frac{\partial \theta}{\partial \Theta} & r \frac{\partial \theta}{\partial Z} \\
    \frac{\partial z}{\partial R} & \frac{1}{R} \frac{\partial z}{\partial \Theta} & \frac{\partial z}{\partial Z}
    \end{bmatrix}
    = 
    \begin{bmatrix}
    \frac{R}{r} & 0 & 0 \\
    0 & \frac{r}{R} & 0 \\
    0 & 0 & 1
    \end{bmatrix}
\end{equation}
}
Cauchy stress tensor for a Neo-Hookean material is derived as,
{\footnotesize
\begin{equation}
    \mathbf{\sigma} = \mu\mathbf{B} - p\mathbf{I} = \mu\mathbf{F}\mathbf{F}^T - p\mathbf{I} = \mu
    \begin{bmatrix}
    \left(\frac{R}{r}\right)^2 & 0 & 0 \\
    0 & \left(\frac{r}{R}\right)^2 & 0 \\
    0 & 0 & 1
    \end{bmatrix}
    - 
    \begin{bmatrix}
    p & 0 & 0 \\
    0 & p & 0 \\
    0 & 0 & p
    \end{bmatrix}
\end{equation}
}
Note that $p$ is a multiplier which imposes incompressibility. So the stress components are obtained as,
{\footnotesize
\begin{equation}
    \begin{aligned}
        \sigma_{rr} &= \mu\left(\frac{R}{r}\right)^2 - p \\
        \sigma_{\theta\theta} &= \mu\left(\frac{r}{R}\right)^2 - p \\
        \sigma_{zz} &= \mu - p
    \end{aligned}
\end{equation}
}
Equilibrium equations in a cylindrical coordinate system are defined as,
{\footnotesize
\begin{equation}
    \begin{aligned}
        &\frac{\partial \sigma_{rr}}{\partial r}+\frac{1}{r}\frac{\partial \sigma_{r\theta}}{\partial\theta}+\frac{\partial \sigma_{rz}}{\partial z}+\frac{1}{r}({\sigma_{rr}-\sigma}_{\theta\theta})+\rho b_r=\rho a_r \\
        &\frac{\partial \sigma_{r\theta}}{\partial r}+\frac{1}{r}\frac{\partial \sigma_{\theta\theta}}{\partial\theta}+\frac{\partial \sigma_{\theta z}}{\partial z}+\frac{2}{r}\sigma_{r\theta}+\rho b_\theta=\rho a_\theta \\
        &\frac{\partial \sigma_{rz}}{\partial r}+\frac{1}{r}\frac{\partial \sigma_{\theta z}}{\partial\theta}+\frac{\partial \sigma_{zz}}{\partial z}+\frac{1}{r}\sigma_{rz}+\rho b_z=\rho a_z \\
    \end{aligned}
\end{equation}
}
Boundary conditions of the problem are,
{\footnotesize
\begin{equation}
    \begin{aligned}
        &\text{at}\ r=a: \quad \sigma_{rr}(a) = -P \quad (\text{where} \quad \lambda_a = \frac{a}{A})\\
        &\text{at}\ r=\infty: \quad \sigma_{rr}(\infty) = 0 \quad (\text{where} \quad \lambda = \frac{r}{R} \rightarrow 1) \\
    \end{aligned}
\end{equation}
}
Assuming all body force and acceleration components are zero, and imposing boundary conditions on equilibrium equations, results in following formulas for constant pressure, Cauchy stress and total strain energy, in terms of stretch ($\lambda_a$).

\begin{equation}
    \begin{aligned}
        &P = \frac{\mu}{2}\left(1-\frac{1}{{\lambda_a}^2}\right) + \mu\ln\left(\lambda_a\right) \\
        &\sigma_{rr}\left(a\right)=\frac{\mu}{2}\left(\frac{1}{{\lambda_a}^2}-1\right)+\mu\ln\left(\frac{1}{\lambda_a}\right) \\
        &SE_0=\pi A^2\mu\left({\lambda_a}^2-1\right)\ln\left(\lambda_a\right)
    \end{aligned}
\end{equation}

\begin{figure}[H]
\centering
    \includegraphics[scale = 0.24] {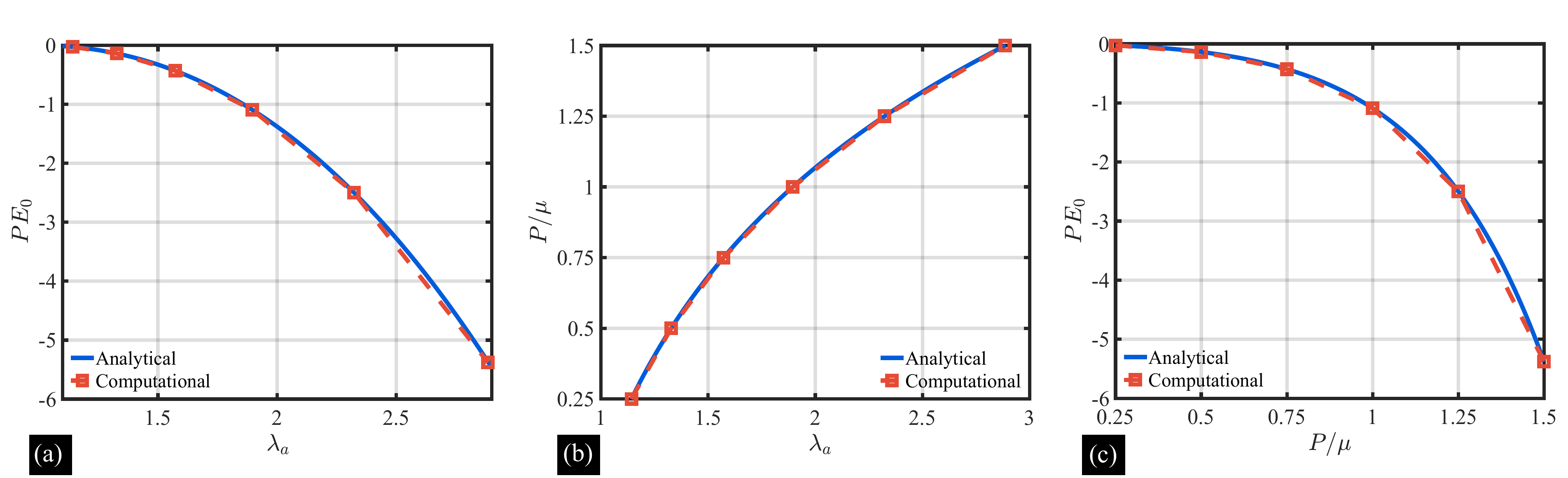}
\caption{Comparison between analytical and computational solutions for single cavity growth in a neo-Hookean medium.}
\label{2D_plots}
\end{figure}

The potential energy of external forces is calculated as $-P\Delta A$,
{\footnotesize
\begin{equation}
    \begin{aligned}
        PEEF_0 = &  -P \pi (a^2 - A^2) = -P \pi A^2 (\lambda_a^2 - 1)
    \end{aligned}
\end{equation}
}
Therefore, the total potential energy of a single cavity in a neo-Hookean medium is
{\footnotesize
\begin{equation}
    PE_0 = SE_0 + PEEF_0 = \pi A^2\mu\left({\lambda_a}^2-1\right)\ln\left(\lambda_a\right) - P \pi A^2 (\lambda_a^2 - 1)
    \label{eq:NH_PE0}
\end{equation}
}
The comparison between the analytical solutions as equation \eqref{eq:NH_PE0}, and the computational solution can be seen in Figure \ref{2D_plots}. The observed consistency verifies the integrity of the computational framework, thereby providing confidence for further extension to the analysis for a pair of cavities.
The codes developed for this research will be shared upon request.

\section{Mesh Convergence Study}
\begin{figure}[H]
\centering
    {
    \includegraphics[scale = 0.25] {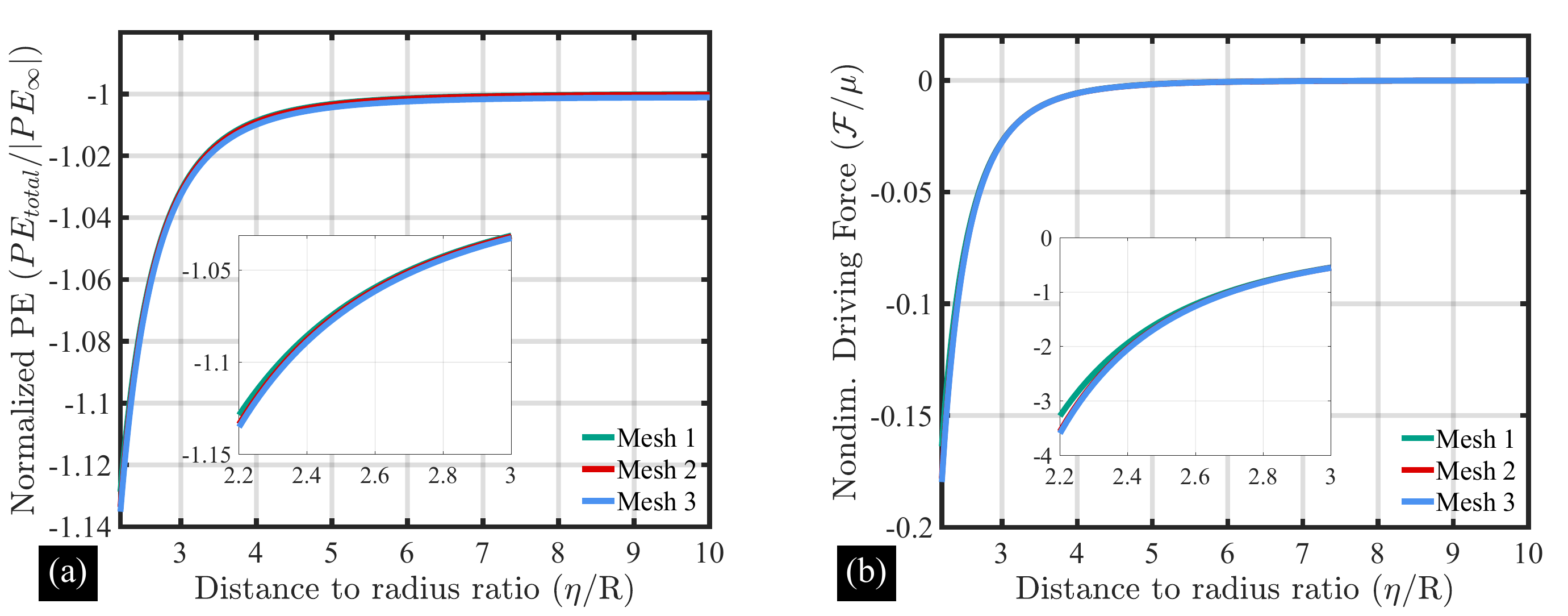}
    \label{meshindep}
    }
    {
    \includegraphics[scale = 0.25] {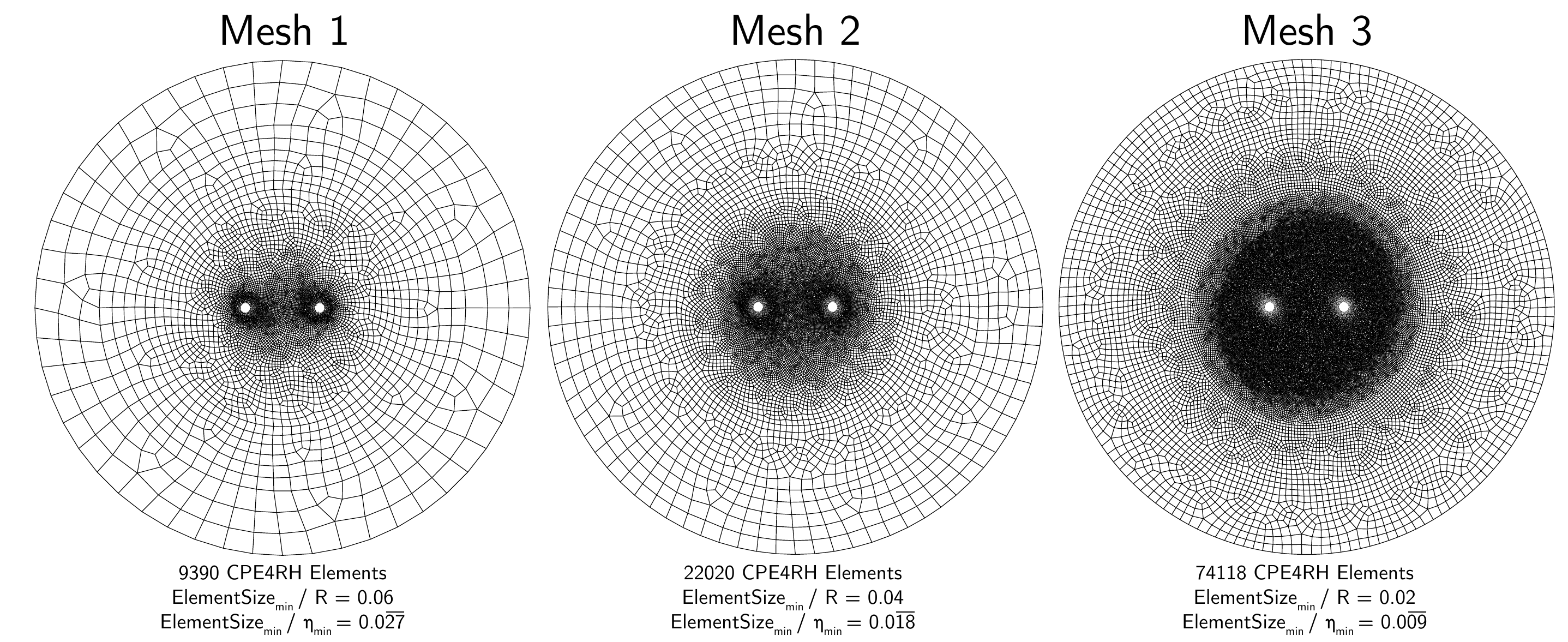}
    \label{MeshOptions}
    }
\caption{Refinement levels ($\eta/R=15$) and solution convergence ($P/\mu=1$).}
\label{convergence}
\end{figure}

\noindent To ensure mesh-independence of the computational solution, we tested three meshes with multiple levels of refinement for a neo-Hookean model, and the result is shown in Figure \ref{convergence}. 
The mesh independence was found to hold across pressures and models.
While mesh refinement primarily influences the results at short-range distances, maintaining mesh independence through the use of Mesh 3 ensures result accuracy.

\section{Deformed Area}
\begin{figure}[H]
\centering
    \includegraphics[scale = 0.25]{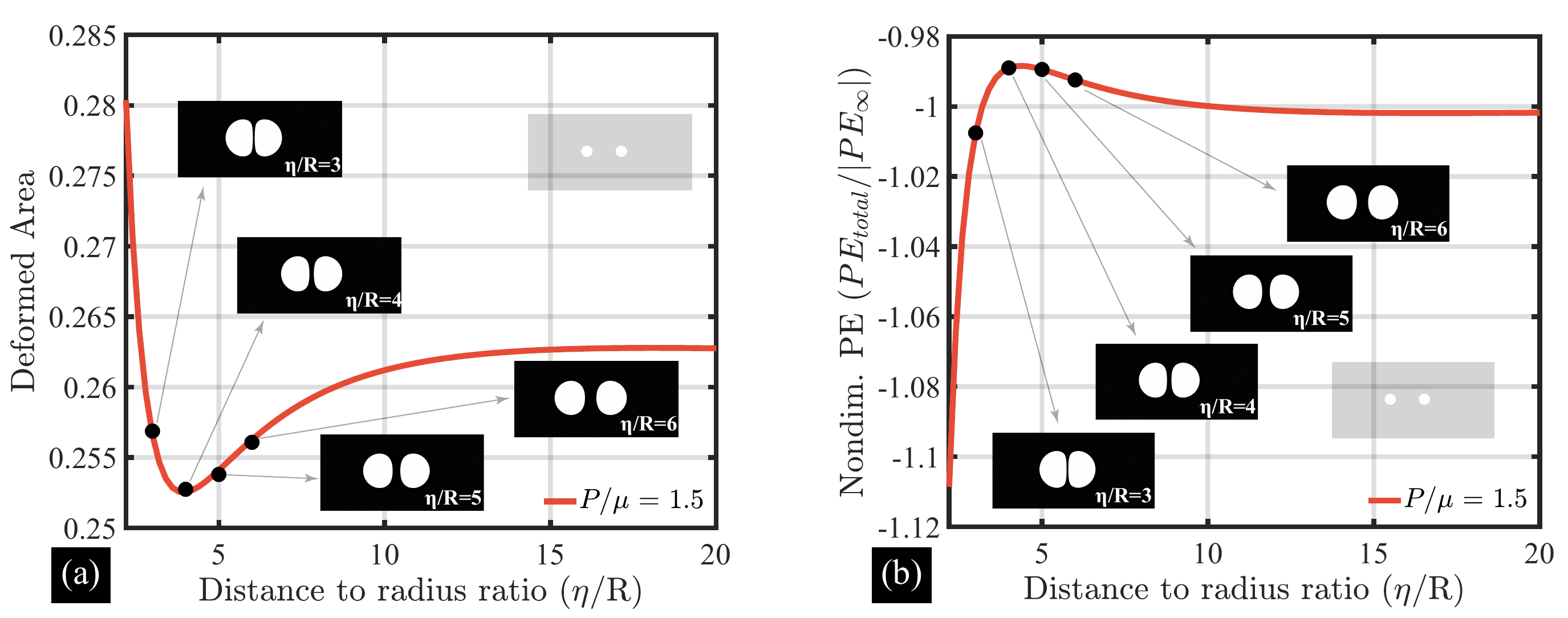}
\caption{\small Area of deformed cavity in neo-Hookean material for $P/\mu = 1.5$.}
\label{deformed_area}
\end{figure}

\noindent Figure \ref{deformed_area} shows the variation of the deformed area as a function of the separation between the cavities for $P/\mu = 1.5$.

\section{Initial Potential Energy Data}
\begin{figure}[H]
\centering
    \includegraphics[scale = 0.24]{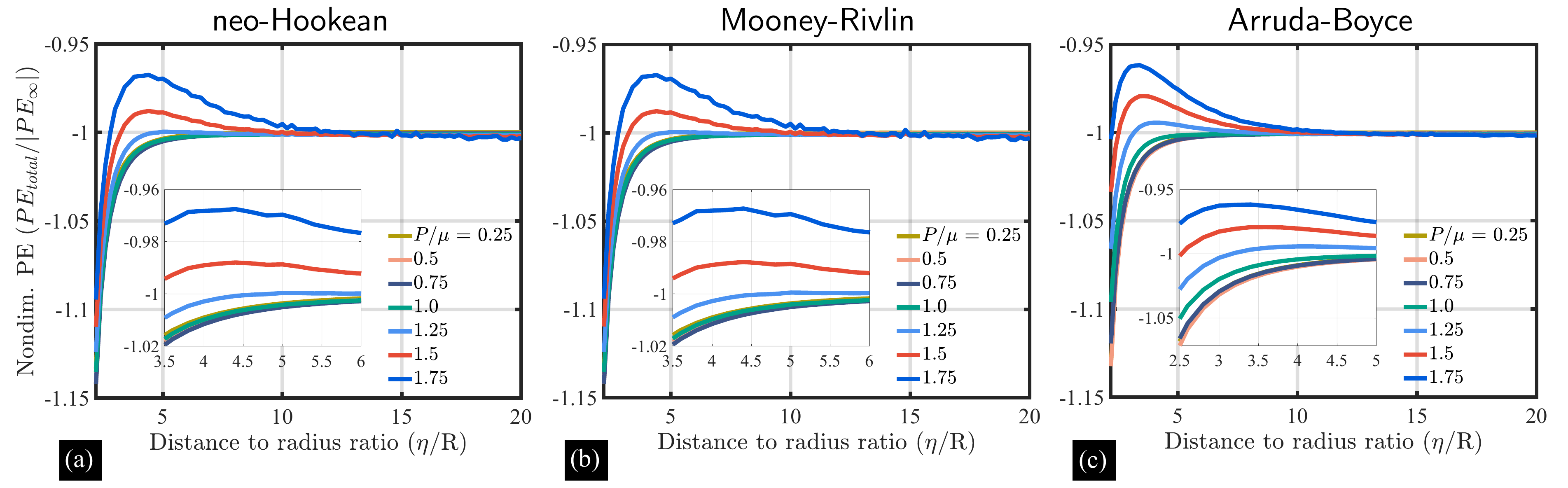}
\caption{\small Original potential energy data prior to fitting.}
\label{original_PE}
\end{figure}

\noindent The initial potential energy data prior to the curve-fitting is presented in Figure \ref{original_PE}. 

 \bibliographystyle{elsarticle-num}
 \bibliography{cas-refs}





\end{document}